\documentclass[12pt]{article}

\input{epsf}
\usepackage[dvips]{graphicx}
\usepackage{cite}
\def\1ad{\mbox{\normalsize $^1$}}
\def\2ad{\mbox{\normalsize $^2$}}
\def\3ad{\mbox{\normalsize $^3$}}
\def\4ad{\mbox{\normalsize $^4$}}

\def\5ad{\mbox{\normalsize $^5$}}
\def\6ad{\mbox{\normalsize $^6$}}
\def\7ad{\mbox{\normalsize $^7$}}
\def\8ad{\mbox{\normalsize $^8$}}

\def\dj{\hbox{d\kern-0.347em \vrule width 0.3em height 1.252ex depth
-1.21ex \kern 0.051em}}

\def\bq{\overline q}
\def\bnu{\overline {\nu}}
\def\nuone{\nu^{(1)}}
\def\bnuone{\bnu^{(1)}}
\def\bone{B^{(1)}}
\def\erone{e^{(1)}_R}
\def\berone{\overline{e}^{(1)}_R}
\def\elone{e^{(1)}_L}
\def\belone{\overline{e}^{(1)}_L}
\newcommand{\ra}{\rightarrow}

\newcommand{\be}{\begin{equation}}
\newcommand{\ee}{\end{equation}}
\newcommand{\ben}{\begin{equation*}}
\newcommand{\een}{\end{equation*}}
\newcommand{\ba}{\begin{eqnarray}}
\newcommand{\ea}{\end{eqnarray}}
\newcommand{\ban}{\begin{eqnarray*}}
\newcommand{\ean}{\end{eqnarray*}}
\newcommand{\brr}{\begin{array}}
\newcommand{\err}{\end{array}}
\newcommand{\bc}{\begin{center}}
\newcommand{\ec}{\end{center}}

\newcommand{\bea}{\begin{eqnarray}}
\newcommand{\eea}{\end{eqnarray}}
\newcommand{\bean}{\begin{eqnarray*}}
\newcommand{\eean}{\end{eqnarray*}}

\newcommand{\ie}{\mbox{\it i.e.~}}

\newcommand\lsim{\mathrel{\rlap{\lower4pt\hbox{\hskip1pt$\sim$}}
    \raise1pt\hbox{$<$}}}
\newcommand\gsim{\mathrel{\rlap{\lower4pt\hbox{\hskip1pt$\sim$}}
    \raise1pt\hbox{$>$}}}

\begin{document} 

\setcounter{page}{0}
\thispagestyle{empty}

%%%%%%%%%%%%%%%%%%%%%%%%%%%%%%%%%%%%%%%%%%%%%%%%%%%%%%%%%%%%%%%%%%%%%%%%%%%%%%%
\begin{flushright}
ANL-HEP-PR-02-032\\
EFI-02-74\\
hep-ph/0206071
\end{flushright}

\vskip 8pt

\begin{center}
{\bf \Large {Is the Lightest Kaluza--Klein Particle a Viable \\[0.25cm]
Dark Matter Candidate?}}
\end{center}

\vskip 10pt

\begin{center}
{\large G\'eraldine Servant $^{a,b}$ and Tim M.P. Tait $^{a}$}
\end{center}

\vskip 20pt

\begin{center}
\centerline{$^{a}$ {\it High Energy Physics Division, Argonne National 
Laboratory, Argonne, IL 60439.}}
\vskip 3pt
\centerline{$^{b}${\it Enrico Fermi Institute, University of Chicago, Chicago, 
IL 60637}}
\vskip .3cm
\centerline{\tt  servant@theory.uchicago.edu, tait@hep.anl.gov}
\end{center}

\vskip 13pt

\begin{abstract}
\vskip 3pt

\noindent
In models with universal extra dimensions (\ie in which all Standard Model 
fields, including fermions, propagate into compact extra dimensions)
momentum conservation in the extra dimensions leads to the conservation of 
Kaluza--Klein (KK) number at each vertex. KK number is violated by loop 
effects because of the orbifold imposed to reproduce the chiral Standard 
Model with zero modes, however, a KK parity remains at any order in 
perturbation theory which leads to the existence of a stable lightest KK 
particle (LKP). In addition, the degeneracy in the KK spectrum is lifted 
by radiative corrections so that all other KK particles eventually decay 
into the LKP.  We investigate cases where the Standard Model lives in
five or six dimensions with compactification radius of TeV$^{-1}$ size 
and the LKP is the first massive state in the KK tower of either the 
photon or the neutrino.  We derive the relic density of the LKP under a
variety of assumptions about the spectrum of first tier KK modes.
We find that both the KK photon and the KK neutrino, with masses
at the TeV scale, may have appropriate annihilation cross sections to
account for the dark matter, $\Omega_M  \sim 0.3$.
\end{abstract}

\vskip 13pt
\newpage
%%%%%%%%%%%%%%%%%%%%%%%%%%%%%%%%%%%%%%%%%%%%%%%%%%%%%%%%%%%%%%%%%%%%%%%%%%%%%
\section{Introduction}

One of the most exciting open questions on the interface between particle 
physics and cosmology is the nature of the dark matter.  In fact, observations
indicate that most of the matter in the universe is dark, and cosmological 
evidence has accumulated to provide independent confirmations that a 
large part of the Dark Matter (DM) is non-baryonic. Recent measurements of 
the Cosmic Microwave Background anisotropy combined with measurement of the 
Hubble parameter suggest a flat universe in which $30 \%$ of the energy 
density is due to non relativistic matter and only $4\%$ is due to baryons, 
consistent with measurements from clusters and Big Bang Nucleosynthesis
(see \cite{Primack:2000xk,Turner:2001mw} for recent reviews). In this paper 
we will use the value derived by Turner from combining 
all the current data\cite{Turner:2001mw}:
\bea
\Omega_M &= &0.33\pm 0.035 , \nonumber \\
h&=&0.69\pm 0.06 ,
\label{hvalue}
\eea
in which $\Omega_M$ is the matter density of the universe expressed as a 
fraction of the critical density for a flat universe. $h$ is the normalized 
expansion rate ($H_0=100 \ h$ km s$^{-1}$Mpc$^{-1}$).

Observational evidence for DM has been building but we still have no solid 
clue as to its identity.  Various candidates have been suggested and the 
theory of structure formation provides indirect evidence about some of its 
properties, strongly hinting that it is weakly interacting and 
non-relativistic at late times. In other words, it is cold dark matter (CDM).
The standard model (SM) of particle interactions, while describing
remarkably well the results of collider experiments, does not contain
a suitable dark matter candidate, and thus it is necessary to consider
extensions.
There are essentially two well motivated DM candidates in this context: 
WIMPs and axions.  WIMPs (Weakly interacting massive particles) were in 
thermal equilibrium with the Standard Model particles in the early universe. 
With masses in the 10--1000 GeV range and weak scale cross sections
($\sigma \sim 10^{-9}$ GeV$^{-2}$) they would have fallen out of 
equilibrium such that their relic density today would correspond to 
$\Omega_{\mbox{\tiny{WIMP}}}\sim {\cal O}(1)$.  Axions,
originally postulated to address the strong CP problem, would not have 
been produced at thermal equilibrium (but through the decay of axionic 
strings or domain walls for instance). Their mass is constrained by 
astrophysical and cosmological arguments to lie in the range 
$m\sim 10^{-5}$--$10^{-2}$ eV.

The most extensively studied DM candidate is the LSP 
(Lightest Supersymmetric Particle), a stable particle in supersymmetric 
(SUSY) models with conserved R-parity, which, in most SUSY scenarios, 
is the {\it neutralino} and is a typical WIMP. A broad range of experiments 
around the world are underway for detecting WIMPs, both through direct 
WIMP--nuclear scattering experiments and through indirect searches such as
detection of cosmic flux  from dark matter annihilation in the 
galactic center. Current searches  are already exploring the parameter space 
of SUSY WIMPs. Unfortunately, SUSY models lack predictability.  They contain 
a huge number of free parameters and one has to make several assumptions 
to reduce this number, for example by assuming a model to describe how
supersymmetry is broken and how the effects of the supersymmetry breaking
are communicated to the superpartners of the SM fields. 
For example, predictions for the cosmic flux from annihilation of the LSP 
in the center of the galaxy can vary over orders of magnitude when 
scanning SUSY parameter space.

While the LSP is very well theoretically motivated, 
since the identity of the DM particles remains unconfirmed we should examine
alternative possibilities. On the other hand, the DM issue sets 
important constraints on model building in particle theory. For any 
extension of the Standard model predicting the existence of a stable particle,
one should compute its cosmological relic density to check whether it 
naturally accounts for DM or if it leads to overclosure of the universe 
in which case the model, or at least the cosmological picture associated 
with it, has to be revised.

The issue when searching for a dark matter candidate is to find a 
{\it stable} particle.  There are two options: 1) The particle essentially 
does not interact with the Standard Model particles, has a very small 
decay rate, and therefore is stable on cosmological scales. 2) The particle
is coupled to the SM. In this case, there must be a symmetry to guarantee 
its stability. In the case of the LSP, there is R--parity to guarantee the 
stability.  In this paper, we study a new DM candidate: the
LKP (Lightest Kaluza--Klein Particle\footnote{The first discussion on the 
relic density of stable Kaluza--Klein particles (referred to as ``pyrgons") 
was made by Kolb and Slansky \cite{Kolb:fm}. It was later 
alluded to in \cite{Dienes:1998vg}.})
 which interacts with SM 
particles and is stable because of a Kaluza--Klein parity. The LKP arises 
in a generic class of models in which all fields propagate in 
extra dimensions. The next section is devoted to explain these 
models, and the particle physics context.  The model has many attractive
features, including the fact that a relatively small number of parameters
are sufficient to describe the LKP. Essentially one: its mass, which at 
tree level 
is the inverse of the compactification radius.  In Section~\ref{sec:relic} 
we review the standard relic density computation. Our major work has been to 
calculate annihilation (and coannihilation) cross sections  for the LKP
in two cases.  In the first case, the LKP is a Kaluza-Klein photon 
(Section~\ref{sec:b1}), while in the second, it is a 
Kaluza--Klein neutrino (Section~\ref{sec:nu1}). We also study the effect of 
coannihilation in Section~\ref{sec:coanni}.  Finally, Section~\ref{sec:sum} 
summarizes our results and discusses open questions which stimulate 
further work on the subject.  Technical details are presented in the 
appendices.  
 
\section{Universal Extra Dimensions}
\label{sec:ued}

Universal extra dimensions (UED) postulate that all of the SM fields
may propagate in one or more compact extra dimensions 
\cite{Appelquist:2000nn}. This is to be contrasted with both the brane 
world scenario \cite{Arkani-Hamed:1998rs} where the SM fields are 
constrained to live in three spatial dimensions while gravity can 
propagate in the bulk, and intermediate models \cite{Antoniadis:1992fh} 
in which only gauge bosons and Higgs fields propagate in extra dimensions 
while fermions live at fixed points. However, there is significant phenomenological
motivation to having fermions and gauge bosons living in the bulk,
including motivation for three families from anomaly cancellation
\cite{Dobrescu:2001ae}, 
attractive dynamical electroweak symmetry breaking (EWSB)
\cite{Cheng:1999bg,Arkani-Hamed:2000hv,He:2001fz},
(supersymmetric) models in which the Higgs mass is a
calculable quantity \cite{Barbieri:2000vh},
preventing rapid
proton decay from non-renormalizable operators 
\cite{Appelquist:2001mj,Arkani-Hamed:1999dc},
orbifold breaking of the parity in left-right
symmetric models \cite{Mohapatra:2002rn}, 
and (through the mechanism of fermion localization)
natural explanations for the observed fermion masses and mixings \cite{Arkani-Hamed:1999dc,Mirabelli:1999ks,Dvali:2000ha,Kaplan:2001ga,Appelquist:2002ft}.
In this article we discover a new motivation for the UED scenario:
to provide a viable dark matter candidate. 

The new feature of the UED scenario compared to the brane world is that
since there is no brane to violate translation invariance along the
extra dimensions, momentum is conserved at tree level leading to
degenerate KK mode masses at each level and
conservation of KK number in the interactions of the four dimensional 
effective theory.  
This statement is broken at the loop level, where 
the fact that the extra dimensions are compact leads to (calculable) 
violations of the full Lorentz symmetry \cite{Hsinchia}, and as a result 
shifts the masses of the KK modes away from their tree level values.

Further violations result by applying orbifold boundary conditions
in order to remove unwanted fermionic degrees of freedom.  These lead to
loop contributions that are log divergent \cite{Georgi:2000ks} 
in the effective theory, thus signalling that they cannot be computed 
but must instead be treated as inputs.  They further correct the KK mode 
masses and break conservation of KK number to conservation of KK 
parity\footnote{KK parity can be seen as the combination 
of a translation by $\pi R$ with a flip of sign of all odd states in the KK Fourier
decomposition of the bulk fields.},
provided the terms induced at both of the orbifold fixed points are
equal.  Whether this will be true or not depends on the details of the
compactification dynamics and the UV completion of the theory, but
the assumption is self-consistent in the sense that if it is true at one scale, the
cut-off scale for instance, it remains true at any scales 
since radiative corrections induce equal terms
on both boundaries.
The resulting theory has interactions only between even numbers of the 
odd-number KK modes.  This conservation of KK parity implies that
the lightest first level KK mode (LKP) cannot decay into 
SM zero modes and 
will be stable, in analogy
with the lightest super-partner in a supersymmetric theory which conserves
$R$-parity.  Thus, UED is the
first extra dimensional scenario to predict a candidate particle for
dark matter.  A further consequence of KK parity, that KK modes must
be pair-produced, leads to interesting collider phenomenology
\cite{Appelquist:2000nn,Rizzo:2001sd}.

At tree level the KK particles of a given level are predicted to be
degenerate with masses $n/R$ where $R$ is the size of the compact
dimension and $n$ is the mode number.  However, at loop level there are 
both calculable and incalculable corrections 
\cite{Hsinchia,vonGersdorff:2002as}.  
We follow the perspective of Ref.~ \cite{Hsinchia} and treat the 
divergent corrections as perturbations on the $1/R$ masses of the KK modes.
This assumption is self-consistent though not completely general, and
could occur, for example, if for some reason the underlying theory
causes them to vanish at the cut-off scale 
($\Lambda \lsim 50 R^{-1}$).  This prescription 
was employed in Ref.~\cite{Hsinchia} and results in small (loop-suppressed)
corrections to the KK mass spectrum induced by renormalization group
evolution from $\Lambda$ to $1/R$.
However, we do not strictly wed ourselves to the particular choice of 
the divergent corrections made in \cite{Hsinchia}, but instead allow 
ourselves the freedom to adjust these terms independently in the effective
theory.

For the LKP to be a well-motivated dark matter candidate, it should be
electrically neutral and non-baryonic.  Thus, the most
promising candidates in the UED picture are first level KK modes of the
neutral gauge bosons (analogues of the KK modes of the photon and $Z$), 
and the KK neutrino, $\nuone$.  One could also consider the first KK mode
of the graviton, though this case seems less promising because its
very weak gravitational interactions would imply that it will annihilate
much less efficiently and could easily overclose the universe.  Since
similar incalculable loop corrections render the graviton mass a separate
input of the theory, we may simply consider that the graviton is
heavier than the LKP, such that at the time scales of interest to us
all of the KK gravitons have already decayed into the LKP and zero modes.
Alternately, one could consider a ``deconstructed model'' 
\cite{Arkani-Hamed:2001ca} in which the extra dimension is represented by
a chain of gauge groups and thus there need not be KK modes of the graviton.  
A simple two-site model can successfully reproduce the physics of the 
first level KK modes, and would be sufficient for our purposes.
More convincingly, there is actually an argument for having the KK graviton heavier
than the lightest KK neutral gauge boson. While the KK graviton receives 
negligible radiative
corrections in contrast with the KK states of the SM particles and therefore 
has a
mass equal to $R^{-1}$, one can easily see from Ref.~\cite{Hsinchia} (see next
paragraph) that the 
lightest
state which diagonalizes the $(\bone,W_3^{(1)})$ mass matrix is lighter 
than $R^{-1}$ for $R^{-1} \gsim 800$ GeV. This is because $\bone$ 
receives negative radiative corrections. So even in the absence of boundary
terms, the LKP, while being nearly degenerate with the KK graviton
(the mass difference is less than 0.1 $\%$) is indeed the KK photon 
for $R^{-1} \gsim 800$ GeV. For smaller compactification scales, one
 can always argue that any tiny negative boundary term correction to 
 the $\bone$ mass will make the KK photon lighter\footnote{One might 
 be worried about the late decay of the KK graviton into the
LKP. Indeed we know that unstable TeV relics may be
dangerous if their lifetime exceeds $10^6$ s because of their effects on the primordially
synthesized abundances of light elements. However, this is true for large
relic densities.
In the case of (massless as well as KK mode) gravitons,
we expect a very suppressed relic number density because of their very
weak coupling. On the other hand, theoretical predictions of relic
gravitons are necessarily subject to
large uncertainties. A reliable estimate is difficult to obtain since it
depends on the complicated dynamics of preheating and on the specific
inflationary model considered.}. Our guideline is 
 phenomenology. The only interesting and plausible cases would correspond 
 to the LKP being a neutral weakly interacting particle.
 
In the gauge boson sector,
EWSB induces mixing between the gauge eigenstates, $\bone$ and $W_3^{(1)}$
in analogy with the familiar effect for the zero modes which produces the
photon and $Z$ boson.  Including tree level contributions, EWSB effects,
and radiative corrections to the masses, the mass matrix in the
$(B^{(n)}, W_3^{(n)})$ basis is \cite{Hsinchia},
\bea
\left(
\begin{array}{cc}
\frac{n^2}{R^2} + \frac{1}{4}g_1^2 v^2 
+ \delta M_1^2 & \frac{1}{4}g_1 g_2 v^2 \\
\frac{1}{4}g_1 g_2 v^2 & \frac{n^2}{R^2} + 
\frac{1}{4}g_2^2 v^2 + \delta M_2^2
\end{array}
\right) ,
\eea
where $g_1$ and $g_2$ are the U(1) and SU(2) gauge couplings, respectively,
$v \sim 174$ GeV is Higgs vacuum expectation value (VEV), $R$ is the radius
of the extra dimension, and $\delta M_1^2$ and $\delta M_2^2$ are the
radiative corrections to the $\bone$ and $W^{(1)}$ masses, including
the boundary terms.  In the absence
of the radiative corrections, the mixing between the KK gauge bosons would
be the same as that for the zero modes, and one would have KK modes of
the photon and $Z$ with the same Weinberg angle as the zero modes.  The
radiative corrections will generally disrupt this relationship, and each KK
level will generally have two neutral bosons which are different mixtures
of $\bone$ and $W^{(1)}$.  As explained above, from an effective theory
point of view $\delta M_1^2$ and $\delta M_2^2$ are separate inputs for
the UED theory, but it is self-consistent to imagine that they are small
and the resulting corrections to the tree-level $n/R$ masses are modest.  
Within this framework
one could also imagine that it is natural to expect 
$\delta M_2^2 > \delta M_1^2$ because $g_2 > g_1$ and $\delta M_2^2$
is further enhanced by larger group factors.  For simplicity, we work in the
limit $\delta M_2^2 - \delta M_1^2 \gg g_1 g_2 v^2$, 
so the mixing angle is effectively driven to zero by the large diagonal
entries in the mass matrix\footnote{This is not so different from the
situation in \cite{Hsinchia}, for which $\sin^2 \theta_W^{(1)} \lsim 0.01$ 
for $1/R > 600$ GeV.}.  Within this framework, one expects the LKP
to be well-approximated as entirely $\bone$.  It thus couples to all
SM fermions (and the Higgs) proportionally to their hypercharges with 
coupling $g_1$, and is approximately decoupled from the gauge bosons.

Similar corrections apply to the Kaluza-Klein modes of the fermions,
and generically their masses are also independent parameters of the
theory.  If one follows the prescription that the lightest particles
are those which undergo only the U(1) hypercharge interaction, the
lightest KK fermion would be the right-handed electron, $\erone$.
Note that the subscript ``$R$'' refers to the fact that it is a KK mode 
of the right-handed electron (and thus is an SU(2) singlet) as opposed
to its chirality; it is a massive Dirac fermion with both right- and 
left-handed polarizations.
In Section~\ref{sec:b1}, we consider the case in 
which $\erone$ is 
substantially heavier than $\bone$, and thus irrelevant in terms of 
its relic abundance while in Section~\ref{subsec:coannie_R}, we also consider
 the case in which $\erone$ is only 
slightly heavier than $\bone$, and thus coannihilation effects can be 
significant.
If one relaxes the restriction that the fields which experience the
SU(2) interaction are heavier than those which only experience the
U(1), one could also consider the $W^{(1)}_3$ or the $\nuone$ as the LKP.
We consider the case in which $\nuone$ is the LKP, including
a variety of coannihilation channels, in Section~\ref{sec:nu1}.

%Our discussion above has implicitly assumed one compact extra dimension,
%for five dimensions (5d) in total.  It is relatively easy to generalize our
%results to consider the six dimensional (6d) case.  At tree level, there
%would be two separately conserved KK numbers, corresponding to the
%quantized momenta in each of the compact dimensions.  Assuming the two
%extra dimensions have the same size, this leads to two equal mass first
%tier KK modes for each SM field which do not interact together with
%zero modes.  This tree-level situation will be modified by radiative 
%effects, which will break both KK numbers down to a single KK parity.
%However, within our framework we treat these effects as small, and
%the physics is well approximated by the tree level picture.
%Generalization to numbers of dimensions higher than six requires some care,
%including careful consideration of the number of fermionic degrees of 
%freedom, and the effects of the orbifold boundary conditions.
%{\bf I need to check these statements carefully, but they should be correct.}

We continue to consider the zero mode gauge bosons in terms of their
well-known mass eigenstates, $\gamma$, $Z$ and $W^\pm$, however we
simplify our results by neglecting all EWSB effects, 
which correct our results at most by $v^2 R^2$.  
Thus, we consider all of the SM fermions and gauge bosons
as massless, and include the full content of the Higgs doublet 
(including the would-be Goldstone bosons) as massless
physical degrees of freedom.
This means that we neglect some processes, such as $\bone \bone \rightarrow$ 
zero mode gauge bosons all together, because they are $v^2 R^2$ suppressed
compared to the dominant decay modes.  It also means that we can choose
to describe the neutral zero mode gauge bosons either in the $Z$ $\gamma$
or in the $B^{(0)}$, $W^{(0)}_3$ basis, as is convenient for the problem
at hand.
This approximation will be further motivated below, where we find that the
favored regions of parameter space for dark matter have the mass of the
LKP on the order of 1 TeV, much greater than $v \sim 174$ GeV.

\section{Density of a Cold Relic Particle}
\label{sec:relic}

In this section, we review the standard calculation of the relic 
abundance of a particle species (see \cite{Kolb:vq,Griest:1990kh} 
for more details) denoted ${\cal Z}$ which was at thermal equilibrium in the 
early universe and decoupled when it was non relativistic.
The evolution of its number density $n$ in an expanding universe is governed 
by the Boltzmann equation:
\be
\frac{dn}{dt}+3Hn=-\langle\sigma v\rangle\left(n^2-{n^{eq}}^2\right) ,
\label{BE}
\ee
where $H=(8\pi\rho/3M_{Pl})^{1/2}$ is the expansion rate of the universe, 
$n^{eq}$ the number density at thermal equilibrium and
$\langle\sigma v\rangle$ is the thermally averaged annihilation cross 
section times the relative velocity.
We are eventually concerned by a massive cold dark matter candidate, for 
which the equilibrium density is given by 
the non relativistic limit:
\be
n^{eq}=g\left(\frac{mT}{2\pi}\right)^{3/2}e^{-m/T} ,
\label{exponentialdecrease}
\ee
where $m$ is the mass of the particle species in question. 
The physics of equation (\ref{BE}) is the following:
At early times, when the temperature was higher than the
mass of the particle, the number density was 
$n^{eq}\propto T^3$, ${\cal Z}$ annihilated with its own
anti-particle into lighter states and {\it vice versa}. As the 
temperature decreased below the mass, $n$
dropped exponentially as indicated in (\ref{exponentialdecrease}) 
and the annihilation rate 
$\Gamma=n\langle\sigma v\rangle$ dropped below $H$. 
The ${\cal Z}$ particles can no longer annihilate and 
their density per comoving volume remains fixed. 
The temperature at which the particle 
decouples from the thermal bath is denoted 
$T_F$ ({\it freeze-out temperature}) and roughly corresponds to the time 
when $\Gamma$ is of the same order as $H$. 

Equation (\ref{BE}) can be rewritten in terms of 
the variable $Y=n/s$, $Y^{eq}=n^{eq}/s$ where $s$ is the entropy
$s={2\pi^2 g_* T^3}/{45}$. $g_*$ counts the 
number of relativistic degrees of freedom.
From the conservation of entropy per comoving volume 
($sa^3=$constant) we get $\dot{n}+3Hn=s\dot{Y}$ so that
\be
s\dot{Y}=-\langle\sigma v\rangle s^2\left(Y^2-{Y^{eq}}^2\right) .
\label{intermediatestep}
\ee
We now introduce the variable:
\be
x=\frac{m}{T} .
\ee
In a radiation dominated era,
\be
H^2=\frac{4\pi^3g_*T^4}{45M_{Pl}^2} \ \ , \ \ t=\frac{1}{2H} \ \ 
\rightarrow \ \ \frac{dx}{dt}=H x ,
\ee 
and (\ref{intermediatestep}) reads:
\be
\label{masterequation}
\frac{dY}{dx}=-\frac{\langle\sigma v\rangle}{H x}s\left(Y^2-{Y^{eq}}^2\right) .
\ee
As is well known, $\langle\sigma v\rangle$ is well approximated by 
a non relativistic expansion (obtained by replacing the square of the 
energy in the center of mass frame by $s=4m^2+m^2v^2)$:
\be
\langle\sigma v\rangle= a+b \langle v^2 \rangle 
+ {\cal O}(\langle v^4 \rangle ) \approx a+6 \ b/x .
\ee
We finally rewrite our master equation (\ref{masterequation}) 
in terms of the variable $\Delta=Y-Y^{eq}$:
\be
\label{deltaequation}
\Delta^{\prime}=-{Y^{eq}}^{\prime}-f(x)\Delta(2Y^{eq}+\Delta) ,
\ee
where 
\be
f(x)=\sqrt{\frac{\pi g_*}{45}}\  m \ M_{Pl}\ (a+6 \ b/x) \  x^{-2} .
\ee
A simple analytic solution can be obtained by studying this equation 
in two extreme regimes. At very early times when
$x<<x_F=m/T_F$, $\Delta^{\prime}<<{Y^{eq}}^{\prime}$ and $\Delta$ is given by:
\be
\Delta=-\frac{{Y^{eq}}^{\prime}}{f(x)(2Y^{eq}+\Delta)} .
\ee
At late times, $\Delta \sim Y>>Y^{eq}$ and 
$\Delta^{\prime}>>{Y^{eq}}^{\prime}$ leading to
\be
\Delta^{-2}\Delta^{\prime}=-f(x) .
\label{latetimes}
\ee
Integrating this equation between $x_F$ and $\infty$ 
and using the fact that $\Delta_{x_F}>>\Delta_{\infty}$:
\be
\Delta_{\infty}=\frac{1}{\int_{x_F}^{\infty}f(x)dx}\approx Y_{\infty} .
\ee
We arrive at:
\be
\label{Y_infty}
Y_{\infty}^{-1}=\sqrt{\frac{\pi g_*}{45}}M_{Pl} \ m \ x_F^{-1}(a+3b/x_F) .
\ee
The contribution of the ${\cal Z}$ particle to the energy density of the 
universe is given by
\be
\Omega_{\cal Z}={\rho_{\cal Z}}/{\rho_c} ,
\ee
where $\rho_c$ is the critical density corresponding to a flat universe,
\be 
\rho_c={3H_0^2 \ M_{Pl}^2}/{8\pi}=1.0539 \times 10^{-5} h^2 
\mbox{GeV cm}^{-3} ,
\ee
The value for $h$ is given in equation (\ref{hvalue}). $\rho_{\cal Z}$ is simply given by 
$\rho_{\cal Z}=m_{\cal Z} n_{\cal Z}=m_{\cal Z} s_0 Y_{\infty}$, 
$s_0=2889.2$ cm$^{-3}$ being the entropy today.
Finally, the contribution to $\Omega$ from a given non relativistic species 
of mass $m_{\cal Z}$ is:
\be
\label{Omega}
\Omega_{\cal Z} h^2 \approx \frac{1.04 \times 10^9}
{M_{Pl}}\frac{x_F}{\sqrt{g_*}}\frac{1}{(a+3b/x_F)} ,
\ee
where $g_*$ is evaluated at the freeze-out temperature.
For our cases of interest, we will have freeze-out temperatures
in the region of 50 GeV, for which we use $g_* = 92$.
Note that the mass $m_{\cal Z}$ does not appear explicitly in this 
expression. Its effect 
is hidden in the coefficients $a$ and $b$ (of dimension GeV$^{-2}$) 
as well as $x_F$.
Therefore, all we have to do is to compute the annihilation cross sections, 
expand them 
in the non relativistic limit and extract the coefficients $a$ and $b$. 
We must also determine $x_F$, the freeze-out temperature.

The freeze-out temperature is defined by solving the equation 
\be
\Delta(x_F)=c \ Y^{eq}(x_F) ,
\label{Tf-definition}
\ee
using the expression for $\Delta(x)$ at early times.
$c$ is a constant of order one determined by matching the late-time 
and early-time solutions. It can be chosen empirically by comparing to a 
numerical integration of the Boltzmann equation. Equation 
(\ref{Tf-definition}) leads to
\be
\label{XF}
x_F=\ln \left(c (c+2) \sqrt{\frac{45 }{8}}
\frac{g}{2\pi^3} \frac{m \ M_{Pl} (a+6b/x_F)}{g^{1/2}_* x^{1/2}_F}\right)
\ee
which is solved iteratively. The result does not depend dramatically on 
the precise value of $c$ which we will take to have the usual value $c=1/2$.

\subsection{Including Coannihilation}
\label{sec:coan}

As pointed out in \cite{Griest:1990kh}, the derivation presented 
above needs to be readdressed in the case of coannihilation. Here, we briefly
 summarize the approach presented in \cite{Griest:1990kh} to be followed 
in this case. Such situation occurs when there are particles nearly 
degenerate with the relic ${\cal Z}$ but with masses slightly greater 
than $m_{\cal Z}$.  These extra particles are nearly abundant as 
${\cal Z}$ and if the mass difference is smaller or of the same order as 
the temperature when ${\cal Z}$ freezes out, they are thermally accessible and 
their annihilation will play a major role in determining the relic 
abundance of ${\cal Z}$. Let us label ${\cal Z}_i$, $i=1,...,N$, these 
particles nearly degenerate in mass. ${\cal Z}_1$ is the LKP, ${\cal Z}_2$ 
is the next LKP, {\it etc}. We also denote $X$, $X^{\prime}$ any zero 
mode (SM) particles. Reactions such as 
${\cal Z}_i {\cal Z}_j \leftrightarrow X X^{\prime}$ 
change the ${\cal Z}_i$ densities $n_i$ and determine their
 abundances. Since all ${\cal Z}_{i>1}$ which survive annihilation 
eventually decay into ${\cal Z}_1$, the relevant quantity is the total 
density of ${\cal Z}_i$ particles $n=\sum_{i=1}^N n_i$, and the Boltzmann
equation for $n$ can be rewritten with accurate approximation 
as \cite{Griest:1990kh}:
\be
\frac{dn}{dt}=-3Hn-\langle \sigma_{\mbox{\tiny{eff}}}v \rangle 
(n^2-{n^{eq}}^2) ,
\label{BEcoanni}
\ee
where 
\be
\label{eq:sigmaeff}
\sigma_{\mbox{\tiny{eff}}}=\sum_{ij}^N\sigma_{ij}\frac{g_ig_j}
{g^2_{\mbox{\tiny{eff}}}}(1+\Delta_i)^{3/2}
(1+\Delta_j)^{3/2} e^{-x(\Delta_i+\Delta_j)} .
\ee$
\sigma_{ij}$ is the cross section
for the reaction  ${\cal Z}_i {\cal Z}_j \rightarrow X X^{\prime}$,
$g_i$ is the number of degrees of freedom of ${\cal Z}_i$,
$\Delta_i=(m_i-m_1)/m_1$, and,
\be
g_{\mbox{\tiny{eff}}}=\sum_{i}^N g_i(1+\Delta_i)^{3/2}e^{-x\Delta_i} .
\label{eq:geff}
\ee
Equation (\ref{BEcoanni}) is of the same form as (\ref{BE}) and can be 
solved using similar techniques. The formula for $x_F$ becomes,
\bea
\label{eq:coxf}
x_F=\ln \left(c (c+2) \sqrt{\frac{45 }{8}}
\frac{g_{\mbox{\tiny{eff}}}}{2\pi^3} \frac{m \ M_{Pl} 
(a_{\mbox{\tiny{eff}}}+6 b{\mbox{\tiny{eff}}}/x_F)}
{g^{1/2}_* x^{1/2}_F}\right) ,
\eea 
where $g$ has been replaced by $g_{\mbox{\tiny{eff}}}$ and 
$a$ and $b$ by $a_{\mbox{\tiny{eff}}}$ and $b_{\mbox{\tiny{eff}}}$, 
the coefficients of the Taylor expansion for $\sigma_{\mbox{\tiny{eff}}}$.
The relic abundance now reads:
\be
\Omega_{{\cal Z}_1} h^2=\frac{1.04 \times 10^9 \; \mbox{GeV}^{-1} \; x_F}
{g_*^{1/2}M_{Pl}(I_a+ 3 I_b/x_f)} ,
\ee
with
\be
I_a= x_F \int_{x_F}^{\infty} \, a_{\mbox{\tiny{eff}}} \, x^{-2}
 \ dx \ \ \ \ , \mbox{and} \ \ \ \
I_b= 2 x_F^2 \int_{x_F}^{\infty} \, b_{\mbox{\tiny{eff}}} \, x^{-3} \ dx\\ .
\ee
We now apply this formalism to our dark matter candidates. 
As explained in section II, we will consider both cases 
${\cal Z}_1=\bone$ and ${\cal Z}_1=\nuone$. 
We have computed the annihilation cross sections of the LKP into any zero 
mode (SM) particle. We begin by ignoring coannihilation and focus on
the $\bone$ and $\nuone$ candidates.  Coannihilation effects will be 
considered in section \ref{sec:coanni}.

\section{$\bone$ as the LKP without Coannihilation}
\label{sec:b1}

We now analyze the case in which the LKP is $\bone$, and all other KK
modes are considerably heavier (roughly 10\% or more \cite{Griest:1990kh}), 
so they do not play a significant role in the final relic density of 
the $\bone$.  The relevant cross sections for pairs of $\bone$ to annihilate
have final states into fermions or into Higgs bosons.  In the limit
in which EWSB effects are neglected, there are no channels into vector
bosons.  For simplicity, we neglect the mass splittings between
the LKP and the higher states in the cross sections,
expressions for which may be found in Appendix~\ref{sec:sigmabone}.  

\begin{figure}[th]
\begin{center}
\includegraphics[height=11cm]{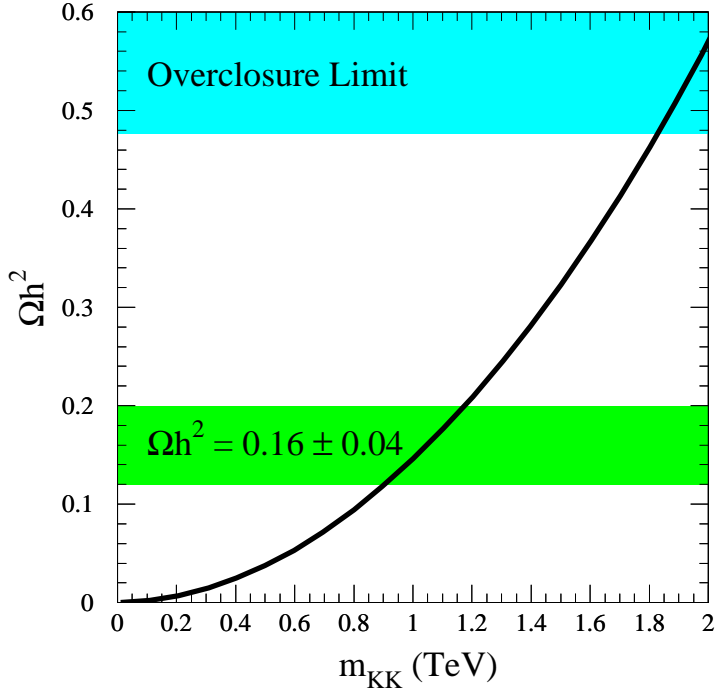}
\end{center}
\caption[]{Prediction for $\Omega_{\bone} h^2$ as a function of the KK
mass (when neglecting coannihilation). The upper horizontal region delimits 
the values of $\Omega h^2$ above which the contribution from ${\bone}$ 
to the energy density would overclose the universe. The lower horizontal
band denotes the region $\Omega =0.33\pm0.035$ (using $h=0.69\pm 0.06$)
and defines the KK mass window if all the dark matter is to be 
accounted for by the $\bone$ LKP.}
\label{fig:omegaphoton}
\end{figure}

From these cross sections, we derive the coefficients in the thermal
average discussed above, finding,
\bea
\nonumber
a =  \frac{4 \pi \alpha_1^2 \left( 2 Y_F + 3 Y_H \right)}{9 m_{KK}^2}, 
\\[0.3cm]
b = -\frac{\pi \alpha_1^2 \left( 2 Y_F + 3 Y_H \right) }{18 m_{KK}^2},
\eea
where $Y_F = 95/18$ and $Y_H = 1/16$. Numerically, for $m_{KK}=1$ TeV, the 
effective annihilation cross section is 
$\sigma \sim 0.6$ pb,
and the annihilation is $35\%$ into quark pairs, $59\%$ into charged
lepton pairs, $4\%$ into neutral leptons, and $2\%$ into Higgs.
For different masses, the cross section falls as $m_{KK}^{-2}$
and the relative importance of the various final states stays
approximately constant.
As discussed above, these results allow us to determine $x_F$, and
we find that it is a very slowly varying function of $m_{KK}$,
decreasing from $x_F=26$ for $m_{KK}=200$ GeV to $x_F=24$ for $m_{KK}=2$ TeV.
Therefore, it is essentially $a$ and $b$ which control the 
$m_{KK}$ dependence of $\Omega_{\bone}$.

In Figure~\ref{fig:omegaphoton} we present the
prediction for $\Omega_{\bone} h^2$ as a function of the KK mass for
five dimensions.  In five dimensions, an upper bound 
$m_{KK} \lsim 1.9$ TeV is set from the universe overclosure condition and 
to account for the dark matter ($\Omega=0.33\pm0.035$), we find that the 
KK mass must lie in the range $m_{KK}\sim 900-1200$ GeV, with a 
corresponding freeze-out 
temperature of order $T_F \sim 36-48$ GeV.  These results are slightly
above the experimental bounds on universal extra dimensions from
precision electroweak data and collider searches ($\sim 350$ GeV for one 
extra dimension \cite{Appelquist:2000nn}),
and imply that provided the fermion KK modes are not very much heavier
than the $\bone$, future collider experiments will be able to study
the region relevant for dark matter.

%In six dimensions, as explained in Section~\ref{sec:ued}, we have two
%species of KK mode of the $B$ (labelled $B^{(1,0)}$ and $B^{(0,1)}$), 
%which we approximate as having no interactions with each other and 
%zero mode fields and we assume are degenerate in mass.  In this case
%our computation of the total relic density proceeds as before, but with
%two (equal) contributions which freeze out at the same temperature.  
%Evident from Figure~\ref{fig:omegaphoton}, the condition not to overclose 
%the universe now requires $m_{KK} \lsim 1$ TeV, and to account for the 
%dark matter we should have $m_{KK} \sim 475-650$ GeV.  These masses are
%again, right at the bound imposed by precision electroweak and collider 
%limits, and indicate that future colliders could be effective
%means to explore the possibility that the dark matter is composed of
%KK modes.

\section{$\nuone$ without Coannihilation}
\label{sec:nu1}

The situation is slightly more intricate in the case where the the 
neutrino is the LKP. To begin with, we now have a relic density composed of 
both the $\nuone$ and its anti-particle, both of which annihilate among 
themselves as well as with each other.  We assume that there is no cosmic
asymmetry between particle and anti-particle in the analysis below.  
If there were a large asymmetry generated before freeze-out, this effect 
could dominate the eventual relic abundance, and the computation below would 
have to be modified.  We must also consider a larger number of annihilation 
processes, including final states of 
fermions, Higgs, and $Z$ and $W^\pm$ gauge bosons.
The various channels are listed in Appendix~\ref{sec:sigmanuone} along
with the necessary cross section formulae.  

We continue to consider the
regime in which the other KK modes are considered light enough that we
can neglect the mass splittings in the cross sections, but heavy enough
that they do not result in a large modification of the final relic density.
One would naturally expect the mass of the $\elone$ to be close to
$\nuone$, its weak partner.  In fact any mass splitting between the
two KK modes is an effect of EWSB, and could lead to dangerously 
large contributions to the $T$ parameter \cite{Peskin:1991sw}.  
Such a contribution could be compensated by,
{\em i.e.} a heavy Higgs boson \cite{Choudhury:2002qb}.  As we will see below
in Section~\ref{subsec:coanninu}, including a degenerate $\elone$ 
will not substantially alter our results.

\subsection{One Flavor}

Our relic density is both $\nuone$ and $\bnuone$
($n_{\cal Z}=n_{\nuone}+n_{\bnuone}$ with $g_{\mbox{\tiny{eff}}}=4$) 
so that the annihilation cross 
section appearing in the Boltzmann equation is 
\bea
\sigma_{\mbox{\tiny{eff}}} & = & \frac{1}{4}\left[
\sigma (\nuone \nuone \ra \nu \nu)+ \sigma(\bnuone \bnuone \ra \bnu \bnu)
+ 2 \ \sigma (\nuone \bnuone \ra X) \right],
\eea
 with 
\bea
 \nonumber
\sigma (\nuone \bnuone \ra X) &=& 
\sigma (\nuone \bnuone \rightarrow q \bq) \ + \ 
\sigma (\nuone \bnuone \rightarrow \nu \bnu) \ + \ 
\sigma (\nuone \bnuone \rightarrow l^+ l^-) \nonumber \\
& & \ + \ \sigma (\nuone \bnuone \rightarrow ZZ)\ + \
\sigma (\nuone \bnuone \rightarrow W^+W^-)  \nonumber \\
& & \ + \ \sigma (\nuone \bnuone \rightarrow \phi \phi^*) ,
\label{leading}
\eea
where the cross section into quarks contains a sum over all quark
flavors, the cross section into neutrinos contains a sum into both
the neutrino zero mode of $\nuone$ and the other flavors, and the
cross section into charged leptons includes both the zero mode charged
partner of $\nuone$, and also the other flavors.  Note that the
matrix elements for annihilation into other flavors are different from those
into zero modes of the same flavor.

Proceeding as before, we expand the effective cross section in 
powers of $1/x_F$, obtaining,
\bea
a_{\mbox{\tiny{eff}}} & = &  
\frac{\alpha^2 \pi \left( 272 s_W^4 - 281 s_W^2 + 154 \right)}
{192 s_W^4 c_W^4 m_{KK}^2} \nonumber \\
b_{\mbox{\tiny{eff}}} & = & -
\frac{\alpha^2 \pi \left( 274 s_W^4 - 127 s_W^2 + 4 \right)}
{1536 s_W^4 c_W^4 m_{KK}^2}
\eea
where the $\nu \bnu$ portion of the result sums over all allowed final states, 
including 3 up- and down-type quarks, 3 charged and neutral leptons, $ZZ$ and
$W^+ W^-$ weak bosons, and the Higgs doublet.  For
$m_{KK} = 1$ TeV, we have the effective cross
section $\sigma_{\mbox{\tiny{eff}}} = 1.3$ pb, slightly higher
than the $\bone$ case. $\sigma_{\mbox{\tiny{eff}}}$ is composed
$18\%$ of the process $\nuone \nuone \ra \nu \nu$, with the remaining
$82\%$ coming from $\nuone \bnuone \ra X$.  This second contribution is
roughly $41\%$ into quarks, $7\%$ / $9\%$ into 
neutral/charged gauge bosons, $2\%$ into Higgs, and $33\%$ / $8\%$
into charged/neutral leptons.

\begin{figure}[t]
\begin{center}
\includegraphics[height=11cm]{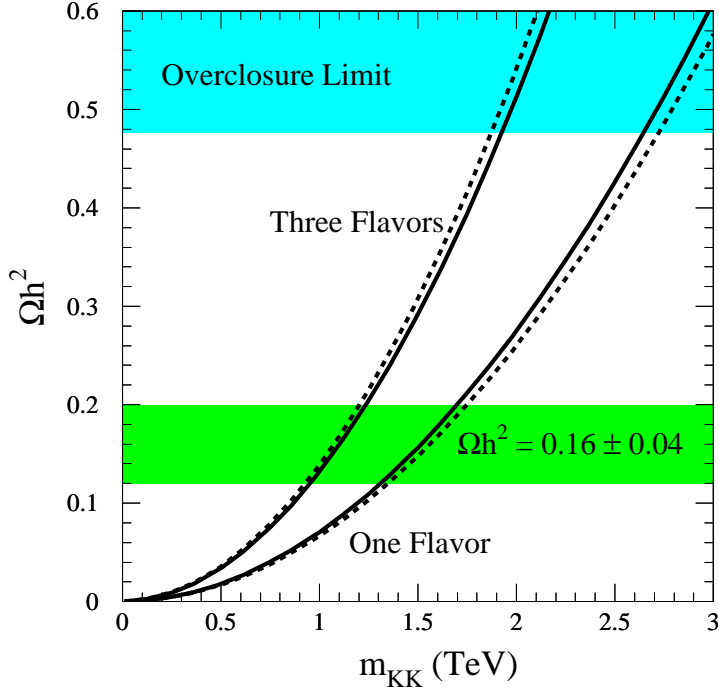}
\end{center}
\caption[]{Prediction for $\Omega_{\nuone} h^2$ as a function of the KK mass.
The solid lines are for $\nuone$ alone (in the one and three family cases) and the dotted ones
correspond to the cases where coannihilation with degenerate $\elone$ is included.}
\label{fig:omeganeutrino}
\end{figure}

Deriving the freeze-out temperature, we find that $x_F $ 
varies from 27 for $m_{KK}=0.2$ TeV to 25 for $m_{KK}=2$ TeV.
$\nuone$ therefore freezes out somewhat later than $\bone$ and thus
has a smaller relic density. The higher effective annihilation cross
section  translates into 
a different prediction for the KK mass to account for the dark matter 
energy density: $m_{KK} \sim 1.3-1.8$ TeV and the overclosure limit is pushed 
up to $2.7$ TeV (see figure \ref{fig:omeganeutrino}).
Again, these values are within the reach of planned experiments such
as the LHC, provided the colored KK mode masses are not significantly
different from the mass of the KK neutrino. 
The value for $x_F$ is not much different from the $\bone$ case, 
however, given the different $m_{KK}$ window, the freeze-out temperature is 
higher, $T_F \sim 50-70$ GeV.

\subsection{Three Flavors}

If we consider three degenerate flavors of KK neutrino, the relic density
is computed as the sum of the densities of all three species plus the
sum of the corresponding anti-particles.  Thus we have 
$g_{\mbox{\tiny{eff}}}=12$.
In this case, the effective cross
section contains three separate contributions identical to that considered
above,  for each species to annihilate among itself, and also additional
cross-flavor channels such as $\nuone_1 \nuone_2 \ra \nu_1 \nu_2$, 
$\nuone_1 \bnuone_2 \ra \nu_1 \bnu_2$, etc
through $t$-channel $Z^{(1)}$ exchange.  There are also cross-flavor 
transitions into charged leptons through $t$-channel $W^{(1)}_\pm$
exchange.  The relevant formulae may be found in 
Appendix~\ref{sec:sigmanuone}.  We continue to assume no 
cosmological asymmetries between particles and anti-particles, and further
consider the case where there are no asymmetries between different flavors.

The effective cross section becomes,
\bea
\sigma_{\mbox{\tiny{eff}}} & = & \frac{1}{12} \left[
\sigma (\nuone_1 \nuone_1 \ra \nu_1 \nu_1) 
+ \sigma (\bnuone_1 \bnuone_1 \ra \bnu_1 \bnu_1)
+ 2 \sigma (\nuone_1 \bnuone_1 \ra X) \right. \nonumber \\
& & \left. \:
\ + \ 2 \sigma (\nuone_1 \nuone_2 \ra \nu_1 \nu_2)
+ 2 \sigma (\bnuone_1 \bnuone_2 \ra \bnu_1 \bnu_2)
+ 4 \sigma (\nuone_1 \bnuone_2 \ra X)
\right]
\eea
where we have assumed that cross sections for all flavors (and combinations
of flavors) are equal.  The cross-flavor annihilation channels are not
as efficient as the same-flavor channels (about twice in size), and
$x_F$ is about the same as in the single flavor case.
Thus, the net result is a larger predicted relic abundance for the 
same mass, as shown in Figure~\ref{fig:omeganeutrino}.  
Thus, the region relevant to explain measurements is lower,
$m_{KK} \sim 950-1250$ GeV, and the overclosure condition requires
$m_{KK} \lsim 1.9$ TeV.  The freeze-out temperature in the relevant
region ranges from $36-47$ GeV.

\section{Coannihilation Results}
\label{sec:coanni}

Coannihilation is expected to play a significant role 
when there are extra degrees of freedom with masses nearly degenerate
with the relic particle.  Experience with the supersymmetric standard
model indicates that large effects are to be expected when
the heavier particles have masses within about $5\%$ of the LSP.
The radiative corrections to the KK spectrum
under the prescription of Ref.~\cite{Hsinchia} indicate that quark and 
gluon KK masses can be shifted by twenty percents. 
Weak gauge bosons also receive corrections (at tree level)
larger than five percents so that the only particles which will be
considered as nearly degenerate with the LKP are the leptons.  We will
simply our analysis by considering all higher Kaluza-Klein modes relevant
for coannihilation to be degenerate, and leave the splitting between
the LKP and next lightest Kaluza-Klein particle (NLKP) as an adjustable
parameter.

As motivated in Section~\ref{sec:ued}, we will compute coannihilation 
channels in the two following situations:
\begin{itemize}
  \item $\bone$ is the LKP and $\erone$ is the NLKP with all other KK modes
        heavy enough that they do not contribute to coannihilation,
  \item $\nuone$ is the LKP and $\elone$ is the NLKP (almost degenerate).
\end{itemize}

The relative mass difference between the LKP and the second LKP is denoted 
by $\Delta = (m_{NLKP} - m_{LKP}) / m_{LKP}$. 

\subsection{$\bone$ Coannihilation with $\erone$}
\label{subsec:coannie_R}

In the first case, we consider $\bone$ as the LKP and 
$\erone$ as the NLKP, assuming no net asymmetry
between the number of $\erone$ and $\berone$.  We consider
both the case with one family of $\erone$, and also the case of
three degenerate families.  With one family, the formula 
(\ref{eq:geff}) for the effective number of degrees of freedom becomes,
\bea
\label{eq:geff1}
g_{\mbox{\tiny{eff}}} & = & 3 + 4 \left( 1 + \Delta \right)^{3/2}
\exp [-x \Delta ] ,
\eea
where we have used $g_f=2,g_{\mbox{\tiny{B$^1$}}}=3$
and summed over $\bone$, $\nuone$, and $\bnuone$.  The effective
annihilation cross section is,
\bea
\nonumber
g^2_{\mbox{\tiny{eff}}} \sigma_{\mbox{\tiny{eff}}} &=& 
g^2_{\mbox{\tiny{B$^1$}}}\sigma( \bone \bone )
+ 4 g_{\mbox{\tiny{B$^1$}}} g_f [1+\Delta]^{3/2} \exp [-x \Delta] 
\sigma (\bone \erone )\\[0.2cm]
& & + 2 g_f^2 [1+\Delta]^3 \exp[-2\Delta x] 
\left( \sigma( \erone \erone ) + \sigma (\erone \berone) \right) ,
\eea
where we have assumed that the cross sections for
annihilation of ($\bone \erone$,$\bone \berone$)
and ($\erone \erone$, $\berone \berone$) into zero modes are equal.

The cross section $\sigma (\bone \bone )$ is as derived before
in Section~\ref{sec:b1}.  The cross section $\sigma( \bone \erone )$
proceeds into final states with zero modes of $e \gamma$ and $e Z$, and in
the limit in which the $Z$ mass is neglected can be equivalently 
described as a single process $\erone \bone \ra e B^{(0)}$.
No $\nu W^-$ final state occurs because $\erone$, being a weak
singlet, does not couple to the SU(2) bosons, and we neglect the
tiny electron Yukawa coupling which would result in a $e \Phi^0$
final state.  There are also channels which convert
$\erone \berone$ into fermions and Higgs through an $s$-channel
$B^{(0)}$ and into two $B^{(0)}$'s (or equivalently, into $ZZ$
$\gamma \gamma$ and $Z \gamma$ final states); and channels in which
$\erone \erone$ exchanges a $\bone$ to become lepton zero modes.
If more than one flavor of $\erone$ has a mass close to $\bone$,
one also has channels in which different flavors of $\erone$
exchange a $t$-channel $\bone$ and thus scatter into their corresponding
zero modes.  All of the needed cross sections are given in 
Appendix~\ref{sec:sigmacoan}.

\begin{figure}[t]
\begin{center}
\includegraphics[height=11cm]{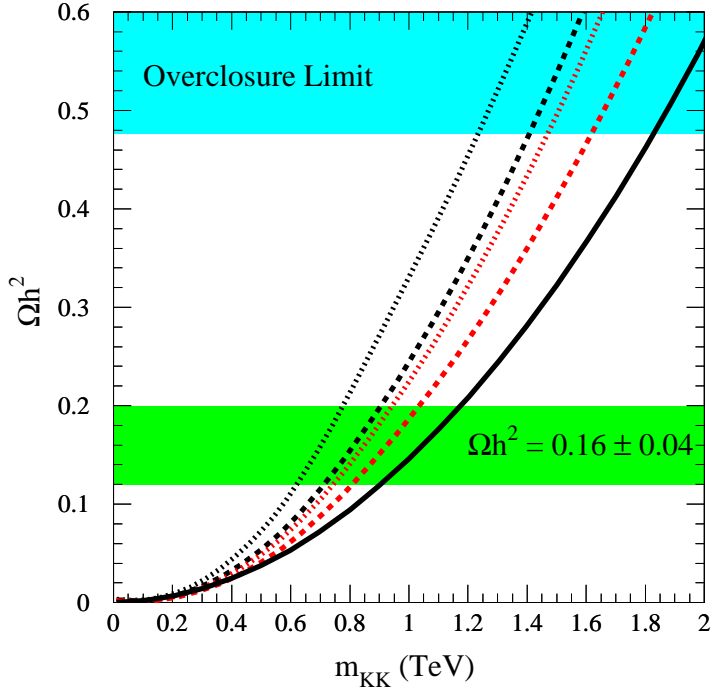}
\end{center}
\caption[]{Prediction for $\Omega_{\bone} h^2$ as in 
Figure~\protect{\ref{fig:omegaphoton}}.  The solid
line is the case for $\bone$ alone, and the dashed and dotted lines correspond
to the case in which there are one (three) flavors of nearly degenerate 
$\erone$.  For each case, the black curves (upper of each pair) 
denote the case $\Delta=0.01$ and the red curves (lower of each pair) 
$\Delta=0.05$.}
\label{fig:omegaber}
\end{figure}

Our result when including $\erone$ almost degenerate with 
$\bone$ ($\Delta=1\%$) is a higher LKP relic density than in the case without 
$\erone$. Indeed, the self annihilation cross section of $\erone$
is not much higher than the one for $\bone$ and the coannihilation cross 
section is significantly smaller (there are only two coannihilation channels 
while $\bone$ and $\erone$ can self annihilate into all zero mode fermions).
This situation is to be contrasted with the SUSY case where coannihilation 
between the neutralino and sfermions can be very efficient and significantly 
reduce the relic density. Here, we have more relics (both $\bone$ and 
$\erone$) which essentially decoupled at the same time 
(and at roughly the same freeze-out temperature as was the
case for $\bone$ alone) and eventually the left over
$\erone$ decay into $\bone$. This translates into a KK mass window 
slightly below the window obtained for $\bone$ alone.  In 
Figure~\ref{fig:omegaber} we present the resulting relic abundance of $\bone$
including both the one flavor and three flavors of $\erone$, for two
choices of $\Delta$ corresponding to $1\%$ and $5\%$ mass splittings.
The curves become approximately degenerate with the $\bone$ without
coannihilation case when $\Delta \gsim 0.1$.  In each case, the
resulting $m_{KK}$ window shifts slightly downward because of the
increase in the predicted relic density, favoring values between
$600-1050$ GeV, depending on the number of light $\erone$ flavors
and the mass splitting.

\subsection{$\nuone$ Coannihilation with $\elone$}
\label{subsec:coanninu}

As mentioned in the introduction of section 5, one should
include $\elone$ in the calculation of the LKP relic density when 
assuming that the LKP is $\nuone$. Indeed, $\nuone$ and 
$\elone$ are expected to be nearly degenerate, with tree level
mass splittings on the order of the mass of the charged lepton.
For one family of leptons we have,
\bea
g_{\mbox{\tiny{eff}}} & = & 4 + 4 \left( 1 + \Delta \right)^{3/2}
\exp [-x \Delta ] ,
\eea 
and the effective annihilation cross section is:
\bea
g_{\mbox{\tiny{eff}}}^2 \sigma_{\mbox{\tiny{eff}}} &=& 
g_f^2 \left\{ 2 \sigma (\nuone \nuone \ra \nu \nu)
           + 2 \sigma (\nuone \bnuone \ra X ) 
\right. \nonumber \\
& & \: \left. \: 
+ \left[ 2 \sigma (\elone \elone \ra e^- e^-) 
+ 2\sigma (\elone \belone \ra X) \right] (1+\Delta)^3 e^{-2\Delta x} \right. 
\nonumber \\
&& \: \left. \: + \left[
4 \sigma ( \nuone \elone \ra \nu e^- )
+ 4 \sigma ( \elone \bnuone \ra X) \right]
(1+\Delta)^{3/2}e^{-x \Delta} \right\}
\eea
where,
\bea
\sigma (\elone \belone \ra X) &=& 
\sigma (\elone \belone \rightarrow q \bq) \ + \ 
\sigma (\elone \belone \rightarrow \nu \bnu) \ + \ 
\sigma (\elone \belone \rightarrow l^+ l^-) \nonumber \\
& & \ + \ \sigma (\elone \belone \rightarrow ZZ,Z \gamma, \gamma \gamma)\ + \
\sigma (\elone \belone \rightarrow W^+W^-)  \nonumber \\
& & \ + \ \sigma (\elone \belone \rightarrow \phi \phi^*) ,
\\
\sigma (\elone \bnuone \ra X )&=&
\sigma (\elone \bnuone \rightarrow q \bq^\prime) \ + \ 
\sigma (\elone \bnuone \rightarrow e^- \bnu) \ + \ \nonumber \\
& & \ + \ \sigma (\elone \bnuone \rightarrow W^- Z)\ + \
\sigma (\elone \bnuone \rightarrow W^+ \gamma)  \nonumber \\
& & \ + \ \sigma (\elone \bnuone \rightarrow \phi \phi^*) .
\eea
For the three family case, we also include the cross flavor annihilation
channels,
\bea
g_{\mbox{\tiny{eff}}} & = &  12 + 12 \left( 1 + \Delta \right)^{3/2}
\exp [-x \Delta ],
\eea
and,
\bea
\hspace*{-.5cm}
g_{\mbox{\tiny{eff}}}^2 \sigma_{\mbox{\tiny{eff}}} &=& 
3 \, g_f^2 \, \left\{ 2 \sigma (\nuone \nuone \ra \nu \nu)
           + 2 \sigma (\nuone \bnuone \ra X )
\right. \nonumber \\
& & \: \left. \: \hspace*{.5cm}
\ + \ 4 \sigma (\nuone_1 \nuone_2 \ra \nu_1 \nu_2 )
+ 4 \sigma (\nuone_1 \bnuone_2 \ra X )
\right. \nonumber \\[0.2cm]
& & \: \left. \: \hspace*{.5cm}
\ + \ \left[ 2 \sigma (\elone \elone \ra e^- e^-) 
+ 2\sigma (\elone \belone \ra X)  \right.  \right. \nonumber \\
&& \: \left. \: \left. \: \hspace*{1.0cm}
\ + \ 4 \sigma (\elone \mu_L^{(1)} \ra e^- \mu^-)
+ 4 \sigma (\belone \mu_L^{(1)} \ra X)
\right] (1+\Delta)^3 e^{-2\Delta x} \right. 
\nonumber \\[0.2cm]
&& \: \left. \: \hspace*{.5cm} \ + \ \left[
4 \sigma ( \nuone \elone \ra \nu e^- )
+ 4 \sigma ( \elone \bnuone \ra X) \right. \right. \nonumber \\
&& \: \left. \: \left. \: \hspace*{1.0cm}
\ + \ 4 \sigma ( \mu_L^{(1)} \bnuone \ra \mu^- \bnu )
+  4 \sigma ( \mu_L^{(1)} \nuone \ra X)
\right]
(1+\Delta)^{3/2}e^{-x \Delta} \right\} ,
\eea
with,
\bea
\sigma (\belone \mu_L^{(1)} \ra X) &=& 
\sigma (\belone \mu_L^{(1)} \rightarrow e^+ \mu^- ) \ + \ 
\sigma (\belone \mu_L^{(1)} \rightarrow \bnu_e \nu_\mu ) , \\
\sigma (\mu_L^{(1)} \nuone \ra X) &=& 
\sigma (\mu_L^{(1)} \nuone \rightarrow \mu^- \nu ) \ + \ 
\sigma (\mu_L^{(1)} \nuone \rightarrow \nu_\mu e^- ) .
\eea
All relevant cross sections are given in Appendix~\ref{sec:sigmacoan}.
From these results, we derive the freeze-out temperature and final relic
density as a function of the mass splitting, 
$\Delta = ( m_{\elone} - m_{\nuone} ) / m_{\nuone}$.  We find that the result
when $\Delta=0$ is a small modification of that with $\nuone$ alone.
The modification of the freeze-out temperature is less than $1\%$ over
the relevant mass range, and the final relic density,
shown in Figure~\ref{fig:omeganeutrino}, is slightly
decreased in the one family case and increased in the three family case.
As $\Delta$ increases, the results rapidly return to the corresponding
$\nuone$ results.

\section{Summary and open questions}
\label{sec:sum}

In this work, we have computed the contribution to the energy density of the 
Universe coming from the relic density of the Lightest Kaluza--Klein 
Particle in two cases: 1) The LKP is a Kaluza--Klein photon, 2) the LKP is 
a Kaluza--Klein neutrino. In models with universal extra dimensions (UED), 
such particle is stable and provides an interesting Dark Matter candidate 
whose mass is the inverse of the compactification radius\footnote{
Note that in models with TeV compactification scale, the higher dimensional Planck
scale is $\sim 10^{14}$ GeV for one extra dimension and $\sim 10^{11}$ GeV
for two extra dimensions.} $R$. One could 
question under which general conditions is the KK parity preserved.
There can indeed be violations of KK parity because of the states
 and interactions
which are localized at the two orbifold fixed points. 
However, if the boundary Lagrangians are symmetric 
under interchange of the two boundaries, KK parity will be preserved and 
unaltered by radiative corrections. The boundary Lagrangian describing 
any coupling 
between states localized at the boundaries and bulk fields is 
invariant under KK parity when fields and couplings are the same at 
the two fixed points. In a string theory context, couplings between bulk modes and
twisted sectors can be predicted. While the symmetry of the boundary 
Lagrangians is not generic, it is not either disfavoured and it is not difficult to
imagine string theories which could preserve KK parity and still have all of the
features necessary. The assumption that KK
parity is preserved  does not conflict decisively with any feature that a low energy
theory derived from string theory must possess.

We have assumed for simplicity that the Standard Model lives 4+1 dimensions. 
One could generalize our results to a higher number of extra 
dimensions\footnote{One should also keep in mind that the computation of 
the relic density is a sensitive function of the expansion rate of the 
universe. The standard calculation assumes that the expansion rate is given 
by the conventional Friedmann equation evaluated in a radiation-dominated 
era corresponding to universe made of a gas of relativistic SM particles. Any 
deviation from this assumption will affect the relic density.}. The 
generalization to the six-dimensional case is reasonably straightforward.
Let us consider for simplicity two extra dimensions with the topology of
a torus, and equal radii.  We impose either a ${\cal Z}_2$ or 
${\cal Z}_4$ orbifold symmetry, so our spaces are $T^2 / {\cal Z}_2$ 
and $T^2 / {\cal Z}_4$.  
For $T^2 / {\cal Z}_2$ our particles have now two KK numbers and there are 
two LKP degenerate in mass: $B^{(1,0)}$ and  $B^{(0,1)}$ and two heavier 
stable states, $B^{(1,1)}$ and $B^{(1,-1)}$ whose tree-level masses are 
about $40\%$ larger.  Following the formalism developed in 
Section~\ref{sec:coan}, we see that for such large mass splittings, we
expect the contribution of the heavier stable states to $\Omega_M$ to be
exponentially suppressed compared to the light states, and thus we
neglect them in the analysis.
Because of KK number conservation (mod 2), at tree level 
the two LKP do not talk to each other and annihilate independently. 
The number of zero modes remains the same, therefore the relic density is 
just twice the one computed in the 5D case. The KK mass window to account 
for the DM is shifted to 650-850 GeV and the limit for overclosure is 
1.3 TeV. 
For $T^2 / {\cal Z}_4$ there are two stable particles with KK numbers 
$(0,1)$ and $(1,1)$.  Again, the heavier state has a tree level 
mass about $40\%$ heavier than the LKP, and we thus conclude that
the prediction for $\Omega_M$ is approximately the same as the 5d case.
The generalization to seven or more dimensions is more subtle 
because the number of fermionic degrees of freedom is modified.
One advantage of this model is that the physics is dominated by a 
single parameter: the size of the extra dimension, $R$.   Interestingly, 
for the UED model to explain the Dark Matter with the LKP, we find that 
$R$ typically has to be of the TeV scale, which is phenomenologically 
interesting and relevant at future colliders.

Having checked that the prediction for $\Omega_M$ is of the right order, 
the next step is to consider detection. Similarly to other WIMPs, 
the direct search for the LKP relies on the deposition of 
$\sim$ keV recoil energy when the WIMP scatters from a nucleus in a detector. 
To study in more details the constraints on the LKP as the dark matter, 
the computation of the corresponding elastic scattering cross section between 
$\bone$ (or $\nuone$) and a nucleus is needed. This task is beyond the 
aim of this paper. In addition, indirect WIMP searches rely on the 
detection of $\gamma$ rays, charged particles or neutrinos from WIMP 
annihilation.  There are two places where annihilation can take place:
\begin{itemize}
\item In the Sun where the LKP may be captured and annihilation greatly enhanced. 
This will 
generate a neutrino spectrum.   The prediction essentially depends on the 
competition between the gravitational capture of the LKP by the Sun
and the LKP annihilation so we would need to know the details of the capture rates and 
of the propagation of the neutrinos from the core to the surface of the Sun to 
make any statement.
\item In the core of the Milky Way. LKP annihilation is important in the 
galactic center where the matter density is higher. To compute the 
resulting spectrum, one needs to know the reprocessing of the direct
products of LKP annihilation. 
\end{itemize}
Among secondary products of annihilation in the galactic center are 
high energy $\gamma$ originating via neutral pion decays (pions result 
from the hadronization of the directly produced quarks) and the synchrotron 
radiation of $e^+e^-$ pairs originating from the decays of charged pions in 
the galactic magnetic field. This requires the implementation of 
fragmentation functions.  However, the flux of neutrinos 
coming from the direct LKP annihilation in the galactic center can be
determined reasonably model-independently. These neutrinos will not be
reprocessed during their journey between the galactic center and us.
To do that, we use a Navarro--Frenk--White profile for the Milky Way of the form: 
\be
\rho_{dm}(r)=\rho_0 \frac{R_0}{r}\,
\left(\frac{ 1 + {R_0}/{a}}{1 + {r}/{a}} \right)^2\,    
\ee
$\rho_0=0.3$ GeV cm$^{-3}$ is the local halo density, $R_0=8$  
kpc the distance between the Sun and the galactic center, $a=20$ kpc 
some length scale. The LKP annihilation rate is then given by
\be
\Gamma=\frac{\sigma v}{m^2}\int^{\infty}_0 \rho^2_{dm}4 \pi r^2 dr
\ee
which leads to a flux of neutrinos at one TeV (assuming a mass of 
1 TeV for the LKP) of 
$4.4 \times 10^{-12}$  cm$^{-2}$ s$^{-1}$ if the LKP is a KK photon,
$2.3 \times 10^{-10}$  cm$^{-2}$ s$^{-1}$ if the LKP is one flavor of 
KK neutrino and $2.8 \times 10^{-9}$  cm$^{-2}$ s$^{-1}$ if the LKP is 
three flavors of KK neutrino.  This is within the sensitivity of 
some future  km$^3$ neutrino telescopes (like IceCube which will reach a sensitivity of 
about $10^{-12}$ in these units at a TeV).
If the photon fluxes are of the same order in first approximation, 
experiments such as MAGIC which will reach about $10^{-12}$ at a TeV 
could set constraints on the mass of the LKP. While a large number of experiments
are underway, this issue of Kaluza--Klein dark matter detection will be worth pursuing. 

\section*{Acknowledgements}

The authors are grateful for discussions with H.-C Cheng, C.-W. Chiang, 
B. Dobrescu, J. Jiang and C.E.M. Wagner. 
We also thank G. Sigl and G. Bertone for 
enlightening us on the cosmic flux issues.  G.S thanks the hospitality of 
the Service de Physique Th\'eorique du CEA Saclay while part of this work was 
being completed.  This work is supported in part by the US Department of 
Energy, High Energy Physics Division, under contract 
W-31-109-Eng-38 and also by the David and Lucile Packard Foundation.

%\newpage
\appendix
\section{$\bone$ Annihilation Cross Sections}
\label{sec:sigmabone}

\begin{table}[t]
\begin{center}
\begin{tabular}{ccc} \hline 
  Initial state & Final state & Feynman diagrams \\ 
 &(zero modes of  SM fields) &  \\ 
 & &\\
$\bone$ $\bone$ &$f$ $\overline{f}$   &  $t(f^{(1)}_L,f^{(1)}_R)$, 
$u(f^{(1)}_L,f^{(1)}_R)$ \\
$\bone$ $\bone$   & $\phi$ $\phi^*$ & $t(\phi^{(1)})$, $u(\phi^{(1)})$,
contact term \\ \hline
\end{tabular}
\end{center}
\caption{Feynman diagrams for which we calculate the 
annihilation cross section of a KK photon into SM particles. 
$s(x)$, $t(x)$ and $u(x)$ denote a tree-level Feynman diagram in which 
particle $x$ is exchanged in the $s$-, $t$- and $u$-channel 
respectively. ``Contact term'' represents the scalar-gauge boson
four point interaction.  $f$ denotes any zero mode fermion and
$\phi$ is the scalar Higgs doublet.}
\label{tab:anniphoton}
\end{table}

In this appendix, we summarize the annihilation cross sections for
$\bone$ into SM fields.  In discussing the various processes it is
worthwhile to remember some general features of the couplings of
the first level KK modes we are considering.  First, KK parity
insures that only vertices with even numbers of first level KK
modes exist.  The two important types are couplings of two KK matter fields
(fermions or Higgs boson)
to a single zero mode gauge boson and coupling of a first level gauge 
boson to a KK matter field and a zero mode matter field.  
Recall that the left- and right-handed components of the zero mode 
fermions each have {\em separate} massive
KK modes, whose couplings to the zero mode gauge bosons are vector-like.
When a KK mode gauge boson couples to a KK fermion and a zero-mode fermion,
there are generally projectors which insure that the zero mode fermion
has the same chirality as the KK fermion.

\begin{figure}[t]
\begin{center}
\includegraphics[height=3cm]{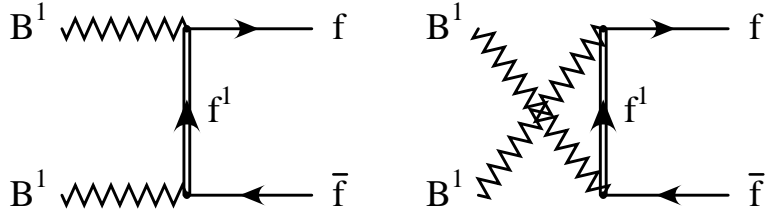}
\end{center}
\caption[]{Feynman diagrams for $\bone \bone$ annihilation into
fermions.}
\label{fig:vvff}
\end{figure}

\begin{figure}[t]
\begin{center}
%\hspace*{-1cm}
\includegraphics[height=3cm]{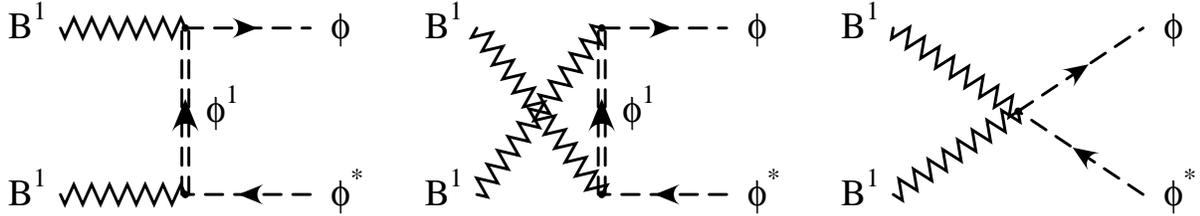}
\end{center}
\caption[]{Feynman diagrams for $\bone \bone$ annihilation into
Higgs scalar bosons.}
\label{fig:vvss}
\end{figure}

In the limit in which electroweak symmetry breaking effects are neglected,
pairs of $\bone$ can annihilate into either zero mode fermions, $f \bar{f}$,
or pairs of Higgs bosons, $\phi \phi^*$.  Figures~\ref{fig:vvff}
and \ref{fig:vvss} show the relevant Feynman diagrams.  
We approximate all first level KK masses ($m$) as equal, and ignore
all ``zero mode'' fermion, scalar and gauge boson masses.
The cross
section for $\bone\bone\rightarrow f \overline{f}$ receives contributions
in which both the $f^{(1)}_L$ and $f^{(1)}_R$ (in the case of
neutrinos only $\nu^{(1)}_L$) are exchanged in both the $t$- and $u$-
channels.  
Summing/averaging over final/initial spins
and integrating over the phase space of the $f \bar f$, the result
may be written,
\bea
\sigma(\bone\bone\rightarrow f \overline{f}) &=& 
\frac{N_c \left( g_L^4 + g_R^4 \right) \,
    \left( 10\,\left( 2\,m^2 + s \right) \,
       {\rm ArcTanh}\left[ \beta \right]  -7 s \beta \right) }
       {72\,\pi \, s^{2}\, \beta^2}
\eea
where $\beta$ is defined as,
\bea
\beta & = & \sqrt{ 1 - \frac{4 m^2}{s} }
\eea
and for each fermion, $g_L = g_1 Y_L$ and $g_R = g_1 Y_R$.  
Note that the massless fermions in the final state have prevented
interference between the graphs in which $f^{(1)}_L$ is exchanged
and $f^{(1)}_R$ is exchanged.
The factor $N_c$ sums over the different color combinations allowed
in the final state, $N_c = 3$ for quarks and $N_c = 1$ for leptons.
Thus, the sum over all three families of SM fermions results in,
\bea
N_c (g_L^4 + g_R^4) & \rightarrow & 3 g^4_1
 (     Y^4_{e_L} + Y^4_{e_R} + Y^4_{\nu_L} 
  + 3( Y^4_{u_L} + Y^4_{u_R} + Y^4_{d_L}  + Y^4_{d_R} )) 
= \frac{95}{18} g_1^4 .
\eea
Annihilation into Higgs, with Feynman diagrams shown in
Fig.~\ref{fig:vvss}, proceeds through $t$- and $u$- channel exchange
of $\phi^{(1)}$ as well as the four-point interaction of 
$\bone \bone \phi \phi^*$.  The cross section is,
\bea
\sigma(\bone\bone\rightarrow \phi \phi^*) & = &
    \frac{g_1^4 Y^4_{\phi}}{6 \pi \beta s}
\eea
where $Y_\phi = 1/2$ and a factor of two is included from the sum over 
the two complex fields in the scalar doublet. 
Note that this result implicitly includes
the decay (after EWSB effects are properly taken into account) into
longitudinal $W$ and $Z$ zero modes as well as the Higgs particle $h$,
as all of these degrees of freedom are included in the scalar doublet.
Neglecting EWSB there are no tree-level decays into gauge bosons;
such decays are induced by EWSB, but are suppressed by powers of
$v^2 / m^2$ and thus we neglect them here.

\section{$\nuone$ Annihilation Cross Sections}
\label{sec:sigmanuone}

The properties of the KK mode of the (left-handed) neutrino are assumed
to be approximately independent of the neutrino species.  For
simplicity, we do not
consider the possibility of a sterile neutrino or its KK modes.
For the purposes of this discussion, we assume a neutrino which
is the weak partner of the left-handed electron; the results
for the weak partners of the muon or tau are simply obtained by
appropriately replacing the exchanged particles in specific processes.
We continue to neglect fermion and boson masses, and ignore fermion
mixing.

\begin{table}[t]
\begin{center}
\begin{tabular}{ccc} \hline 
  Initial state & Final state & Feynman diagrams \\ 
 & (zero modes of  SM fields)&  \\ &&\\
$\nuone$ $\bnuone$ & $q$ $\bq$       & $s(Z)$\\
                   & $\phi$ $\phi^*$ & $s(Z)$ \\
                   & $\nu$ $\bnu$    & $s(Z)$, $t(\bone,W_3^{(1)})$\\
                   & $e^-$ $e^+$     & $s(Z)$, $t({W_+}^{(1)})$\\
                   & $Z^0$ $Z^0$     & $t(\nuone)$, $u(\nuone)$\\
                   & $W^- W^+$       & $t(e^{(1)}_L)$, $s(Z)$ \\  
$\nuone$ $\nuone$  & $\nu$ $\nu$     & $t(\bone,W_3^{(1)})$, 
                                       $u(\bone,W_3^{(1)})$ \\
$\nuone_1$ $\nuone_2$ & $\nu_1$ $\nu_2$ & $t(Z^{(1)})$\\
\hline
\end{tabular}
\end{center}
\caption{Same as Table \ref{tab:anniphoton} but for annihilation of 
the KK neutrino.}
\label{tab:annineutrino}
\end{table}

\begin{figure}[t]
\begin{center}
\includegraphics[height=3cm]{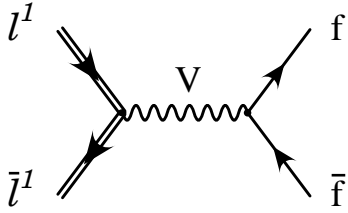}
\end{center}
\caption[]{Feynman diagrams for $\nuone \bnuone$ annihilation into
quarks or leptons of other families.}
\label{fig:llff}
\end{figure}

The $\nuone$ can annihilate with $\bnuone$ into quark (and other
family lepton) zero modes through
an $s$-channel $Z$ zero mode (Figure~\ref{fig:llff}).
The cross section is given by,
\bea
\sigma (\nuone \bnuone \rightarrow f \overline{f}) & = &
\frac{N_c g_Z^2 \, \left( \bar{g}^2_L + \bar{g}^2_R \right)
    \left( s + 2 \, m^2 \right) }
{24 \, \pi \,\beta \,s^2} ,
\label{eq:sigmannff}
\eea
where,
\bea
g_Z &=& \frac{e}{ 2 s_W c_W} ,
\eea
are the couplings of the $Z^0$ to $\nuone \bnuone$ and
$\bar{g}_{L(R)}$ are the standard zero mode couplings between the $Z^0$
and $f \bar f$,
\bea
\label{eq:gbar}
\bar{g}_{L(R)} = \frac{e}{s_W c_W} \left[ T^3 - Q_f s_W^2 \right]
\eea
where $T^3$ is the third component of weak iso-spin of $f$ and $Q_f$ its 
charge.  $N_c$ accounts for the sum over final state color configurations,
as before.

\begin{figure}[t]
\begin{center}
\includegraphics[height=3cm]{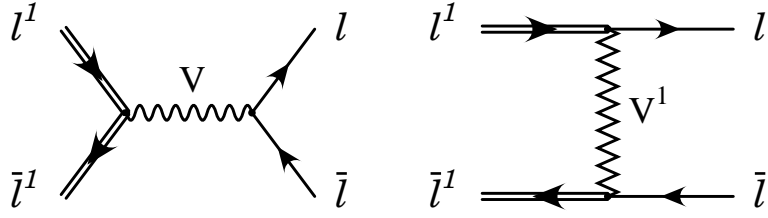}
\end{center}
\caption[]{Feynman diagrams for $\nuone \bnuone$ annihilation into
zero mode leptons of the same family.}
\label{fig:llnn}
\end{figure}

Annihilation into zero modes of the charged lepton partner $e^+ e^-$ 
proceeds either through an $s$-channel $Z$ or a $t$-channel $W_{+}^{(1)}$
(Figure~\ref{fig:llnn}), or into its own zero modes 
($\nu \overline{\nu}$) through
an $s$-channel $Z$ zero mode or by exchanging a $t$-channel
$W_3^{(1)}$ or $\bone$ (Figure~\ref{fig:llnn}).  In the limit in
which we ignore the mass splitting between $W_3^{(1)}$ and $\bone$,
the exchange of both can be summed into a single $Z^{(1)}$ exchange.
The cross section into either zero mode neutrinos of the same flavor
or their charged lepton weak partners is,
\bea
\sigma ( \nuone \bnuone \rightarrow \ell \overline{\ell}) & = & 
\frac{
g_Z \hat{g}^2_L \overline{g}_L
[5 \beta s + 2 (2 s + 3 m^2) L ] }
{32\, \pi \, \beta^2 \,s^2} 
+ \frac{g_Z^2 (\overline{g}^2_L + \overline{g}^2_R )
\,\left( s + 2 \, m^2 \right)}
{24 \, \pi \, \beta \,s^2}
\nonumber \\
& & + 
\frac{\hat{g}^4_L \left[ \beta \, ( 4 s + 9 m^2 )
+ 8 m^2 L \right]}
{64 \, \pi \, m^2 \, \beta^2 \,s} 
\label{eq:sigmalblb}
\eea
where $\hat{g}_L = g_Z$,
\bea
L & = & \log \left[ \frac{ 1 - \beta }{ 1 + \beta } \right] ,
\eea
and $\overline{g}_{L(R)}$ are given as before.
For annihilation into charged leptons, the $\overline{g}_{L(R)}$ should be
replaced by the charged lepton values from Eq.~\ref{eq:gbar} and
$\hat{g}_L$ now corresponds to the $W_\pm^{(1)}$ coupling to $\nuone$ and
$e^0$,
\bea
\hat{g}^e_L & = & \frac{e}{\sqrt{2} s_W} .
\eea

\begin{figure}[t]
\begin{center}
\includegraphics[height=3cm]{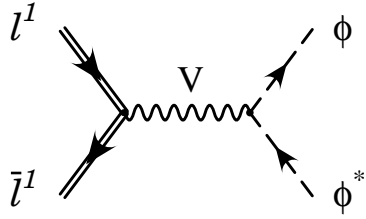}
\end{center}
\caption[]{Feynman diagrams for $\nuone \bnuone$ annihilation into
scalar Higgs bosons.}
\label{fig:llss}
\end{figure}

First modes of $\nuone \bnuone$ can annihilate into zero modes of Higgs bosons
through an $s$-channel $Z$ zero mode (Figure~\ref{fig:llss})
with cross section,
\bea
\sigma ( \nuone \bnuone \rightarrow \phi_i \phi_i^*) & = &
\frac{g^2_\phi g_Z^2 \left( s + 2 m^2 \right)}
{48 \,\pi \, \beta \, s^2} ,
\label{eq:sigmannss}
\eea
where the $g_Z$ coupling of $\nuone$ to a $Z$ zero mode is as before,
and there are two couplings of $\phi \phi^*$ to $Z^0$ from the charged
and neutral entries of the doublet, respectively,
\bea
g_\phi & = & \frac{e}{s_W c_W}
\left( T^3 - Q_\phi s_W^2 \right) ,
\label{eq:zphicoupling}
\eea
where the first (upper entry in the doublet) Higgs has charge $Q_\phi=1$ and
the second has $Q=0$.

\begin{figure}[t]
\begin{center}
\includegraphics[height=3cm]{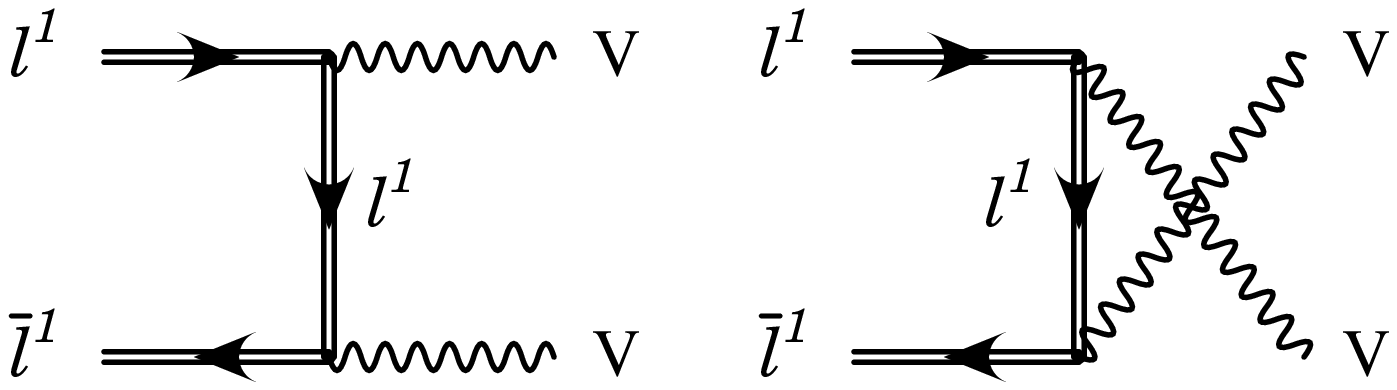}
\end{center}
\caption[]{Feynman diagrams for $\nuone \bnuone$ annihilation into
two neutral vector bosons $V V$.}
\label{fig:llzz}
\end{figure}

Annihilation into $Z Z$ are mediated by $t$- and $u$-channel $\nuone$,
(Figure~\ref{fig:llzz}) and has cross section,
\bea
\sigma ( \nuone \bnuone \rightarrow Z Z) & = &
\frac{g_Z^4 \left( 
2 \left[ s^2 + 4 m^2 s - 8 m^4 \right] {\rm ArcTanh}[ \beta ]
- \beta s [s + 4 m^2] \right) }
{8\,\pi \, \beta^2 \,s^3 }
\label{eq:sigmannzz}
\eea
where $g_Z$ is the coupling to the zero mode $Z$ boson defined above.

\begin{figure}[t]
\begin{center}
\includegraphics[height=3cm]{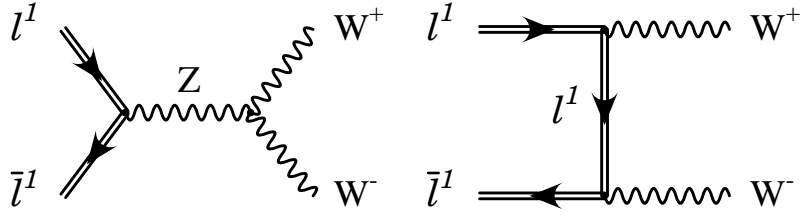}
\end{center}
\caption[]{Feynman diagrams for $\nuone \bnuone$ annihilation into
two charged vector bosons $W^+ W^-$.}
\label{fig:llww}
\end{figure}

Annihilation into $W^+ W^-$ zero modes includes $t$-channel
$e^{(1)}_L$ exchange and $s$-channel annihilation through a
virtual $Z$ zero mode (Figure~\ref{fig:llww}).  The cross section is,
\bea
\sigma ( \nuone \bnuone \rightarrow W^+ W^-) & = &
\frac{-5 g_{ZWW}^2 g_Z^2 [s + 2 m^2 ]}{24 \pi \beta s^2} 
+ \frac{g_W^2 g_{ZWW} g_Z [ \beta s - 2 m^2 L ]}{8 \pi \beta^2 s^2}
\nonumber \\
& & 
- \frac{g_W^4 [ \beta (s + 4 m^2) + (s + 2 m^2) L ]}{8 \pi \beta^2 s^2}
\label{eq:sigmannww}
\eea
where $g_Z$ are the couplings to the $Z$ zero mode as before,
$g_W$ is the (vector-like) coupling between the $W$ zero mode to $e_L^{(1)}$ 
and $\nuone$,
\bea
g_W & = & \frac{e}{\sqrt{2} s_W} ,
\eea
and $g_{ZWW}$ is the $Z$-$W^+$-$W^-$ coupling between zero modes,
\bea
g_{ZWW} & = & e \frac{c_W}{s_W} .
\eea

\begin{figure}[t]
\begin{center}
\includegraphics[height=3cm]{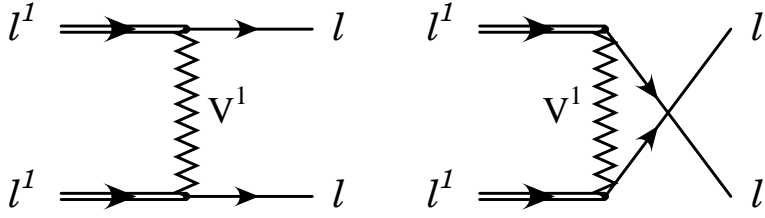}
\end{center}
\caption[]{Feynman diagrams for $\nuone \nuone$ annihilation into
two zero mode leptons $\nu \nu$.}
\label{fig:nnnn}
\end{figure}

Furthermore,
$\nuone \nuone$ ($\bnuone \bnuone$) can annihilate into 
$\nu \nu$ ($\overline{\nu} \overline{\nu}$) through $t$- and
$u$-channel exchange of $W_3^{(1)}$ or $\bone$ (Figure~\ref{fig:nnnn}).
The cross section for annihilation of two neutrinos 
$\nuone \nuone \rightarrow \nu \nu$ is given by,
\bea
\sigma (\nuone \nuone \rightarrow \nu \nu) & = &
\frac{\hat{g}_L^4
( \beta s (2 s - m^2) + 2 m^2 (4 s - 5 m^2) {\rm ArcTanh} [ \beta ] )}
{32 \pi \beta^2 s^2 m^2}
\label{eq:sigmallll}
\eea
where $\hat{g}_L$ is defined above.
Note that if more than one KK neutrino species is present, two
related processes will also take two neutrinos of different species
into their two zero modes, or one neutrino and
one anti-neutrino of different flavors into their zero modes.  
In both cases, we have a single $t$-channel Feynman diagram and the 
cross sections are,
\bea
\sigma (\nuone_1 \nuone_2 \rightarrow \nu_1 \nu_2) & = &
\frac{\hat{g}_L^4 ( 4 s  -3 m^2 ) }
{64 \pi \beta s m^2} , 
\label{eq:sigman1n2} \\
\sigma (\nuone_1 \bnuone_2 \rightarrow \nu_1 \bar{\nu}_2) & = &
\frac{\hat{g}_L^4 \left( \beta (4 s + 9 m^2) + 8 m^2 L \right) }
{64 \pi \beta^2 s m^2} ,
\label{eq:sigman1n2b}
\eea
with $\hat{g}_L = g_Z$.  Finally, we can have cross-flavor
transition between $\nuone_1 \bnuone_2$ into charged lepton
zero modes, $e^-_1 e^+_2$ through a $t$-channel $W^{(1)}_\pm$ 
exchange.  The corresponding cross section is given by the result
for $\sigma (\nuone_1 \bnuone_2 \rightarrow \nu_1 \bar{\nu}_2)$
with $\hat{g}_L = e / \sqrt{2} s_W$.

\section{Coannihilation Cross Sections}
\label{sec:sigmacoan}

\begin{table}[t]
\begin{center}
\begin{tabular}{ccc} \hline 
  Initial state & Final state & Feynman diagrams \\ 
 & (zero modes of  SM fields)&  \\ &&\\
$\erone$ $\berone$ & $q$ $\bq$         & $s(B^{(0)})$\\
                   & $\nu$ $\bnu$      & $s(B^{(0)})$\\
                   & $e^-$ $e^+$       & $s(B^{(0)})$, $t(\bone)$\\
                   & $\phi$ $\phi^*$   & $s(B^{(0)})$ \\
                   & $Z$ $Z$           & $t(\erone)$, $u(\erone)$\\
                   & $\gamma$ $\gamma$ & $t(\erone)$, $u(\erone)$\\
                   & $Z$ $\gamma$      & $t(\erone)$, $u(\erone)$\\
$\erone$ $\erone$  & $e^-$ $e^-$       & $t(\bone)$, $u(\bone)$ \\
$\erone$ $\bone$   & $e^-$ $\gamma$    & $s(e^-)$, $t(\erone)$ \\
                   & $e^-$ $Z$         & $s(e^-)$, $t(\erone)$ \\
\hline
\end{tabular}
\end{center}
\caption{Same as Table \ref{tab:anniphoton} but for coannihilation of
$\erone$.}
\label{tab:annier}
\end{table}

The $\erone$ can annihilate with $\berone$ into quark (and other
family lepton) zero modes through an $s$-channel $B^{(0)}$ 
(Figure~\ref{fig:llff}).
The cross section is given by,
\bea
\sigma (\erone \berone \rightarrow f \overline{f}) & = &
\frac{N_c g_1^4 Y^2_{e_R}  \left( Y^2_{f_L} + Y^2_{f_R} \right)
    \left( s + 2 \, m^2 \right) }
{24 \, \pi \,\beta \,s^2} .
\label{eq:sigmaeeff}
\eea
Note that since $\erone$ is a weak singlet, this formula
also applies for annihilation into zero modes of the neutral lepton partner.

Annihilation into $e^+ e^-$ zero modes proceeds
through an $s$-channel $B^{(0)}$ or by exchanging a $t$-channel
$\bone$ (Figure~\ref{fig:llnn}).  The cross section is,
\bea
\sigma ( \erone \berone \rightarrow e^+ e^-) & = & 
\frac{ g_1^4 Y^4_{e_R}
[5 \beta s + 2 (2 s + 3 m^2) L ] }
{32\, \pi \, \beta^2 \,s^2}
+ \frac{g_1^4 Y^4_{e_R} \left[ \beta \, ( 4 s + 9 m^2 )+ 8 m^2 L \right]}
{64 \, \pi \, m^2 \, \beta^2 \,s}
\nonumber \\ & &
+ \frac{ g_1^4 Y^2_{e_R} \left( Y^2_{e_R} + Y^2_{e_L} \right)
\,\left( s + 2 \, m^2 \right)}
{24 \, \pi \, \beta \,s^2}
\label{eq:sigmallee}
\eea
where, $L$ is defined in Section~\ref{sec:sigmanuone}.
Note that we have chosen to
include the decay into left-handed electrons in this result, though
we could have equally well considered it part of Eq.~(\ref{eq:sigmaeeff}).
Evident from comparison of the two equations, this is simply a 
matter of book-keeping.

Annihilation into zero modes of Higgs bosons occurs through an $s$-channel 
$B^{(0)}$ (Figure~\ref{fig:llss}) with cross section,
\bea
\sigma ( \erone \berone \rightarrow \phi \phi^*) & = &
\frac{ g_1^4 Y^2_{e_R} Y^2_\phi
\left( s + 2 m^2 \right)}
{24 \,\pi \, \beta \, s^2} ,
\label{eq:sigmaeess}
\eea
where the factor of 2 to sum over the two entries of the doublet is included.

Annihilation into $Z Z$, $Z \gamma$, and $\gamma \gamma$ 
are mediated by $t$- and $u$-channel $\erone$ (Figure~\ref{fig:llzz}).
This process can be more conveniently described as the single channel
$\erone \berone \ra B^{(0)} B^{(0)}$, equivalent to the sum of
these three $\gamma$ and $Z$ processes.  The cross section is,
\bea
\hspace*{-1cm}
\sigma ( \erone \berone \rightarrow B^{(0)} B^{(0)} ) & = &
\frac{g_1^4 Y^4_{e_R} \left( 
2 \left[ s^2 + 4 m^2 s - 8 m^4 \right] {\rm ArcTanh}[ \beta ]
- \beta s [s + 4 m^2] \right) }
{8\,\pi \, \beta^2 \,s^3 }
\label{eq:sigmaeezz}
\eea
Annihilation into $W^+ W^-$ 
zero modes is zero in the limit in which one ignores EWSB effects, 
because $\erone$ is a singlet.

Finally we have the process $\erone \erone \ra e^- e^-$
( $\berone \berone \ra e^+ e^+$ ) via exchange of a $t$-
or $u$-channel $\bone$.  This cross section is
\bea
\sigma (\erone \erone \rightarrow e^- e^-) & = &
\frac{g_1^4 Y^4_{e_R}
( \beta s (2 s - m^2) + 2 m^2 (4 s - 5 m^2) {\rm ArcTanh} [ \beta ] )}
{32 \pi \beta^2 s^2 m^2} ,
\label{eq:sigmaeeee}
\eea
for lepton KK modes of the same flavor, and,
\bea
\sigma (\erone \mu^{(1)}_R \rightarrow e^- \mu^-) & = &
\frac{g_1^4 Y^4_{e_R} ( 4 s  -3 m^2 ) }{64 \pi \beta s m^2} \\
\sigma (\erone \overline{\mu}^{(1)}_R \rightarrow e^- \mu^+) & = &
\frac{g_1^4 Y^4_{e_R} \left( \beta ( 4 s  + 9 m^2 ) + 8 m^2 L \right) }
{64 \pi \beta^2 s m^2}
\eea
for two modes of different lepton flavor.

\begin{figure}[t]
\begin{center}
\includegraphics[height=3cm]{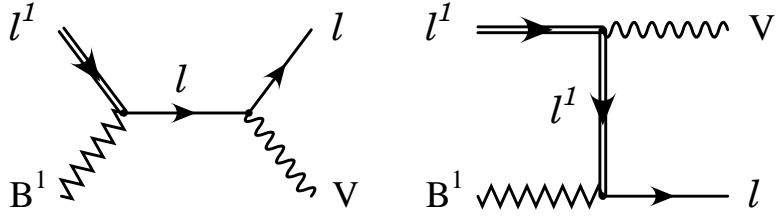}
\end{center}
\caption[]{Feynman diagrams for $\bone f^{(1)}$ annihilation into
a zero mode $f$ and vector boson.}
\label{fig:bnvn}
\end{figure}

Coannihilation of a $\bone$ with $\erone$ (or $\berone$) proceeds into
either $e^- \, Z$ or $e^- \, \gamma$.  In the limit in which the $Z$ mass is 
disregarded, we can equally well describe this as a single channel into
$e B^{(0)}$.  The cross section is given by,
\bea
\sigma (\bone \erone \rightarrow B^{(0)} e^-) & = &
\frac{Y_{e_R}^4 g_1^4 [ \beta s (2 m^2 - s)+ m^2 (6 m^2 - 2 s) L ]}
{48 \pi \beta^2 s^2 m^2}
\nonumber \\
& &
+
\frac{Y_{e_R}^4 g_1^4 (s - m^2)}{96 \pi \beta s m^2}
\nonumber \\
& &
+
\frac{Y_{e_R}^4 g_1^4 [\beta s (s - 4 m^2) - 2 m^2 (s+ 6 m^2) L]}
{96 \pi \beta^2 s^2 m^2}
\label{eq:sigmaeb}
\eea

\begin{table}[t]
\begin{center}
\begin{tabular}{ccc} \hline 
  Initial state & Final state & Feynman diagrams \\ 
 & (zero modes of  SM fields)&  \\ &&\\
$\elone$ $\belone$ & $q$ $\bq$         & $s(\gamma,Z)$\\
                   & $\nu$ $\bnu$      & $s(Z)$, $t(W_\pm^{(1)})$\\
                   & $e^-$ $e^+$       & $s(Z,\gamma)$, $t(\bone,W_3^{(1)})$\\
                   & $\phi$ $\phi^*$   & $s(\gamma,Z)$ \\
                   & $Z$ $Z$           & $t(\elone)$, $u(\elone)$\\
                   & $\gamma$ $\gamma$ & $t(\elone)$, $u(\elone)$\\
                   & $\gamma$ $Z$      & $t(\elone)$, $u(\elone)$\\
                   & $W^+$ $W^-$       & $s(\gamma,Z)$, $u(\nuone)$\\
$\elone$ $\elone$  & $e^-$ $e^-$       & $t(\bone,W_3^{(1)})$ \\
$\elone$ $\mu_L^{(1)}$& $e^-$ $\mu^-$  & $t(\bone,W_3^{(1)})$ \\
$\mu_L^{(1)}$ $\belone$& $\mu^-$ $e^+$ & $t(\bone,W_3^{(1)})$ \\
                   & $\nu_\mu$ $\bnu_e$& $t(\bone,W_\pm^{(1)})$ \\
$\elone$ $\bnuone$ & $q$ $\bq^\prime$  & $s(W_-)$ \\
                   & $e^-$ $\bnu$      & $s(W_-)$, $t(\bone,W_3^{(1)})$\\
                   & $\phi$ $\phi^*$   & $s(W_-)$ \\
                   & $Z$ $W^-$         & $s(W_-)$, $t(\elone)$, $u(\nuone)$\\
                   & $\gamma$ $W^-$    & $s(W_-)$, $t(\elone)$, $u(\nuone)$\\
$\elone$ $\nuone$  & $e^-$ $\nu$       & $t(\bone,W_3^{(1)})$, $u(W_-^{(1)})$\\
$\mu_L^{(1)}$ $\nuone$  &  $\mu^- \nu$  & $t(\bone,W_3^{(1)})$ \\
                        &  $\nu_\mu e^-$& $u(W_\pm^{(1)})$ \\
$\mu_L^{(1)}$ $\bnuone$ & $\mu^- \bnu$ & $t(\bone,W_3^{(1)})$\\
\hline
\end{tabular}
\end{center}
\caption{Same as Table \ref{tab:anniphoton} but for coannihilation of
$\elone$.}
\label{tab:anniel}
\end{table}

The KK modes of the left-handed electron are somewhat more complicated
because they involve the SU(2) bosons as well as the U(1) boson.  We find
it convenient to consider neutral zero mode gauge bosons in the 
$W_3^{(0)}, B^{(0)}$ basis.  Annihilation into fermions is given by,
\bea
\sigma (\elone \belone \rightarrow f \overline{f}) & = &
\frac{N_c \: g^4 \left( s + 2 \, m^2 \right) }{24 \, \pi \,\beta \,s^2} .
\label{eq:sigmaeeff2}
\eea
where the coupling,
\bea
g^2 & = & \left( g_1^2 \; Y_{f} \; Y_{e_L} 
+ g_2^2 \; T^3_f \; T^3_{e_L} \right) ,
\eea
is in terms of the hypercharges ($Y$) and third component of weak
iso-spin ($T^3$) for the fermion and the left-handed electron.
For annihilation into zero modes of electrons of the same family we
also have $t$-channel exchange of $\bone$ and $W_3^{(0)}$,
and for annihilation into zero modes of the neutral lepton partner
we have $t$-channel $W_\pm^{(1)}$ exchange.  Both results may be expressed,
\bea
\sigma ( \elone \belone \rightarrow e^+ e^-, \nu \bnu) & = & 
\frac{ \hat{g}^2_L g^2
[5 \beta s + 2 (2 s + 3 m^2) L ] }
{32\, \pi \, \beta^2 \,s^2}
+ \frac{ \hat{g}^4_L \left[ \beta \, ( 4 s + 9 m^2 )+ 8 m^2 L \right]}
{64 \, \pi \, m^2 \, \beta^2 \,s}
\nonumber \\ & &
+ \frac{ g^4
\,\left( s + 2 \, m^2 \right)}
{24 \, \pi \, \beta \,s^2}
\label{eq:sigmallee2}
\eea
where $g^2$ is defined above, for the specific case of $f$ an electron
or neutrino, and $\hat{g}_L = e / 2 s_W c_W$ for electrons and
$\hat{g}_L = e / \sqrt{2} s_W$ for the neutrino final state.

Under our approximation in which EWSB effects are neglected, annihilation 
of $\elone \belone$ into $W$ bosons is equal to the process 
$\nuone \bnuone \ra W^+ W^-$
given in Equation~(\ref{eq:sigmannww}), by SU(2) invariance.  Similarly,
the sum of the annihilation processes into $Z Z$, $\gamma Z$, and
$\gamma \gamma$ are also equal to the process $\nuone \bnuone \ra Z Z$ 
given in Equation~(\ref{eq:sigmannzz}) and the sum of annihilation
into both components of the Higgs doublet is given by
the process $\nuone \bnuone \ra \phi \phi^*$ in 
Equation~(\ref{eq:sigmannss}).  Furthermore, like-sign annihilation
$\elone \elone \ra e^- e^-$ is equal to $\nuone \nuone \ra \nu \nu$,
and thus is given by Equation~(\ref{eq:sigmallll}).

$\elone \bnuone$ annihilate into zero modes of quarks or other family leptons
through an $s$-channel $W^{(0)}_-$ with cross section given
by Equation~(\ref{eq:sigmaeeff2}), replacing 
with the appropriate charged current coupling: $g^2 \ra g_2^2 / 2$. 
Annihilation into leptons of the same family also includes $t$-channel
exchange of $Z^{(1)}$.   The cross section is given
by Equation~(\ref{eq:sigmallee2}) with 
$g^2 \ra g_2^2 / 2, \hat{g}^2_L \ra (e / 2 s_W c_W)^2 [ -1/2 + s_W^2 ]$.
Annihilation into Higgs occurs through an $s$-channel $W_-^{(0)}$ and may be 
obtained from Equation~(\ref{eq:sigmannss}) with the replacement 
$g_Z^2 g_\phi^2 \ra e^4/4 s_W^4$.  Annihilation into gauge bosons is most 
simply described in terms of annihilation into $W^- B^{(0)}$, and 
$W^- W_3^{(0}$.  The first process is mediated by $t$-channel $\elone$ 
exchange and $u$-channel $\nuone$ exchange, and is given by 
Equation~(\ref{eq:sigmannzz}) with $g_Z^2 \ra -e^2 / 2 \sqrt{2} s_W c_W$.  
The second is simply obtained from $\nuone \bnuone \ra W^+ W^-$ 
in Equation~(\ref{eq:sigmannww}) by SU(2) invariance.  Finally, 
$\elone \nuone \ra e^- \nu$ is mediated by $t$-channel 
$Z^{(1)}$ and $u$-channel $W^{(1)}_\pm$ exchange, and is given by,
\bea
\sigma (\elone \nuone \rightarrow e^- \nu) & = &
\frac{g_t^2 g_u^2 \left[ 
4 m^2 (4s - 5 m^2) {\rm ArcTanh} [ \beta ] + m^2 \beta s \right]}
{32 \pi \beta^2 s^2 m^2} \nonumber \\
& & + \: \frac{\beta s \left[ (g_t^4 + g_u^4)(4 s - 3 m^2) \right]}
{64 \pi \beta^2 s^2 m^2} ,
\eea
with $g^2_t = e^2/(2 s_W^2 c_W^2) (-1/2+s_W^2)$ 
and $g^2_u = e^2 / 2 s_W^2$

If there are multiple families of KK left-handed electrons or neutrinos, there
will also be flavor-changing annihilation of 
$\mu_L^{(1)} \nuone \ra \mu^- \nu$, $\mu_L^{(1)} \nuone \ra \nu_\mu e^-$,
and $\mu_L^{(1)} \bnuone \ra \mu^- \bnu$.  The cross sections for the
first two processes are both given by Equation~(\ref{eq:sigman1n2}), with
the replacements $\hat{g}^2_L \ra e^2/(2 s_W^2 c_W^2) (-1/2+s_W^2)$,
and $\hat{g}^2_L \ra  e^2 / 2 s_W^2$.  The cross section for
$\mu_L^{(1)} \bnuone \ra \mu^- \bnu$ is given by 
Equation~(\ref{eq:sigman1n2b}) with the replacement 
$\hat{g}^2_L \ra e^2/(2 s_W^2 c_W^2) (-1/2+s_W^2)$.

%%%%%%%%%%%%%%%%%%%%%%%%%%%%%%%%%%%%%%%%%%%%%%%%%%%%%%
%%%%%%%%%%%%%%%%%%%%%%%%%%%%%%%%%%%%%%%%%%%%%%%%%%%%%

\end{document}